# Over what length scale does an inorganic substrate perturb the structure of a glassy organic semiconductor?

*Kushal Bagchi[a], Chuting Deng[b], Camille Bishop[a], Yuhui Li[c], Nicholas E Jackson[b,d], Lian Yu[a,c], M.F Toney[e], J.J de Pablo[b,d], M.D Ediger[a]*

[a]Department of Chemistry, University of Wisconsin-Madison, Madison, Wisconsin 53706, United States

[b]Pritzker School of Molecular Engineering, University of Chicago, Chicago, Illinois 60637, United States.

[c]School of Pharmacy, University of Wisconsin-Madison, 777 Highland Avenue, Madison, Wisconsin 53705-2222, United States

[d]Center for Molecular Engineering and Materials Science Division, Argonne National Laboratory, Lemont, Illinois 60439, United States, Lemont, Illinois 60439, United States

[e]Stanford Synchrotron Radiation Lightsource, SLAC National Accelerator Laboratory, Menlo Park, California 94025, United States

**Abstract:** While the bulk structure of vapor-deposited glasses has been extensively studied, structure at buried interfaces has received little attention, despite being important for organic electronic applications. To learn about glass structure at buried interfaces, we study the structure of vapor-deposited glasses of the organic semiconductor DSA-Ph (1,4-di-[4-(*N*,*N*-diphenyl)amino]styryl-benzene) as a function of film thickness; structure is probed with grazing incidence X-ray scattering. We deposit on silicon and gold substrates and span a film thickness range of 10-600 nm. Our experiments demonstrate that interfacial molecular packing in vapor-deposited glasses of DSA-Ph is more disordered compared to the bulk. At a deposition temperature near room temperature, we estimate ~ 8 nm near the substrate can have modified molecular packing. Molecular dynamics simulations of a coarse-grained representation of DSA-Ph reveal a similar length scale. In both the simulations and the experiments, deposition temperature controls glass structure beyond this interfacial layer of a few nanometers.

**Introduction:**

Interfaces between different materials or phases are critically important in materials science as they have a strong influence on mechanical, electrical, and optical properties. For molecular materials, understanding and controlling structure at such buried interfaces has been a significant challenge. The broken translational symmetry at a buried interface results in structure and dynamics quite different from that observed in the adjacent bulk materials, and new analytical techniques have been developed to better understand buried interfaces[1,2]. In the last few decades, while great strides have been made in understanding the interfacial structure of molecular crystals[3], liquids[4] and liquid crystals[5] the structure of molecular glasses near solid interfaces has received little attention. In addition to serving as model systems for amorphous materials[6,7], molecular glasses also have important applications. For instance, molecular glasses formed by organic semiconductors are the active layers in OLEDs (organic light emitting diodes) which are used in commercial cell-phone and television displays[8,9] and are explored for use in other technologies[10]. In such organic electronics applications, the structure at the buried interface of an organic semiconductor and an inorganic electrode influences charge injection barriers[11].

Molecular glasses used in OLED applications are prepared by physical vapor-deposition (PVD). Vapor-deposited glasses, unlike liquid-quenched glasses, exhibit structural anisotropy[12,13,14]. The anisotropic structure of 100-1000 nm thick vapor-deposited glasses has been investigated in the last decade and often can be understood using the surface equilibration mechanism[15,16]; for these thick films, the substrate temperature and the deposition rate control the glass structure[17,18]. On the other hand, there is little understanding regarding the interfacial structure of PVD glasses. There is no mechanism or theory that can predict the structure of vapor-deposited glasses near a solid substrate. Understanding the structure of PVD glasses at buried interfaces remains an outstanding challenge in the field of molecular solids, with considerable technological implications. By way of analogy, a number of studies of the interface of crystalline and semi-crystalline organic semiconductors with inorganic substrates has led to structure-property relationships in the context of organic-field effect transistors (OFETs)[19,20]. For OLEDS, in contrast to OFETs, the glassy state is preferred[12] and an understanding of the interface of amorphous organic semiconductors with inorganic layers might similarly be expected to lead to improved device performance.

A starting point towards understanding the interface of a PVD glass with an inorganic solid is to determine over what length scale the substrate can perturb glass structure. In the crystalline state, packing at the buried interface of organic semiconductors has been extensively studied[21,22]. For several molecular semiconductors, such as pentacene, the substrate promotes new crystalline packing arrangements not found in the bulk[3]. Different "substrate-induced phases" can form on different substrates,[3] and for thin films of pentacene, substrate-induced phases can propagate at least 100 nm[23] away from the interface. Presently even the length scale over which the substrate can influence the structure of a vapor-deposited organic glass is poorly understood. While some measurements suggest that the substrate can influence the average structure of PVD glass films as thick as 100 nm[24], other studies suggest that the range of influence is at least an order of magnitude smaller[25]. This discrepancy is important as the typical thickness for an organic glass layer in an OLED device is ~ 30 nm.

In this work, we quantify the length scale over which an inorganic substrate can perturb the structure of PVD glasses of DSA-Ph (1,4-di-[4-(*N*,*N*-diphenyl)amino]styryl-benzene). DSA-Ph is used in OLED devices as a blue-light emitter and the structure of its thick vapor-deposited glasses has been previously characterized with x-ray scattering[26]. We study the structure of DSA-Ph films deposited on Si/$SiO_2$ (down to 13 nm) and Au substrates (down to 25 nm) as a function of film thickness using grazing incidence X-ray scattering. These experiments indicate that interfacial molecular packing in vapor-deposited glasses of DSA-Ph is more disordered compared to the bulk. We also perform computer simulations of vapor-deposited glasses of a coarse-grained representation of DSA-Ph. We find, both in experiments and simulations, that beyond the first 3 to 8 nm, the structure of a vapor-deposited organic glass is independent of the underlying substrate. Beyond this critical length-scale the deposition conditions (substrate temperature and deposition rate) determine the structure of a vapor-deposited glass.

**Experimental Methods:**

**Sample Preparation:** DSA-Ph was purchased from Luminescence Technology Corp (LT-N631); the powder had a purity (HPLC) greater than 99%**.** DSA-Ph was deposited as received without further purification. The samples were deposited in a vacuum chamber with a base pressure of ~ $10^{-6}$ Torr. The deposition rate was monitored in real time using a quartz crystal microbalance (QCM). The deposition rate for all the reported samples was approximately 0.2 nm/s.

**Substrates**: Depositions on silicon were performed on ⟨100⟩ cut wafers. The silicon wafers had ~2 nm of native oxide. For depositions on gold, ~ 10 nm of gold was deposited on a silicon substrate by sputtering using a Leica EM ACE 600 Coater. The gold was deposited at a rate of ~0.15 nm/s with a sputtering current of 30 mA and an argon pressure of $2.5\times10^{-2}$ mbar. An AFM image of the gold substrate is shown in Figure S2.

**Thickness Measurements:** After deposition the thicknesses of the films were measured using variable angle spectroscopic ellipsometry (VASE) using a Woollam M-2000 instrument. Psi (amplitude ratio) and delta (phase difference) were obtained at incidence angles of 50 $^{o}$, 60 $^{o}$ and 70$^{o}$ (the angle between the surface normal and the incident beam). The thickness was obtained from a Cauchy model by fitting data in a wavelength range of 600-1000 nm. All measurements were performed at ambient temperature. For films thinner than 70 nm, optical constants were fixed to those obtained for thicker films, and only thickness was fit. For films thinner than 70 nm, the thickness obtained from VASE, on an average, is about ~ 6% higher than the estimate from the QCM. Based on this comparison we estimate that for films thinner than 70 nm, the thicknesses reported below could, on an average, be systematically higher by ~ 6%; this does not produce any important ambiguity in the interpretation of our results.

**GIWAXS:** GIWAXS measurements were performed in BL 11-3 at SSRL with a photon wavelength of 0.973 Å. All measurements were performed at room temperature. A "$\chi$ correction" was performed to account for the grazing geometry, resulting in the missing wedge along $Q_z$[27] . All reported data is at an angle of incidence of 0.14 $^o$ (the incidence angle in GIWAXS is between the incident beam and the substrate plane; 0 $^o$ at complete grazing and 90 $^o$ in transmission). For order parameter evaluation, data from 1.35 to 1.45 Å$^{-1}$ was summed at each angle in reciprocal space; this region was chosen as there is maximum diffracted intensity from DSA-Ph in this region. Data in the missing wedge ($\chi$ = 0-10°) was obtained by extrapolation. (Here $\chi$ is the azimuthal angle in reciprocal space, with $\chi$= 0 $^o$ defined by $Q_z$.) To evaluate the background contribution, scattered intensity from 0.75 -0.85 Å$^{-1}$ and 1.95-2.05 Å$^{-1}$ was averaged. The exact choice of regions for background subtraction did not influence the observed order parameter; almost the same order parameters were obtained when intensity from 0.9-1.0 Å$^{-1}$ and 1.8-1.9 Å$^{-1}$ was averaged and used for the background subtraction. The background subtracted intensity was used for evaluation of the Hermans order parameter, $S_{GIWAXS}$, using the following equations:

$$S_{GIWAXS} = \frac{1}{2}(3 < \cos^2\chi > -1) \qquad (1)$$

with $<\cos^2 \chi>$ evaluated as follows:

$$< \cos^2 \chi > = \frac{\int_0^{90} \mathrm{I}(\chi)(\cos^2 \chi)(\sin \chi) d\chi}{\int_0^{90} \mathrm{I}(\chi)(\sin \chi) d\chi} \qquad (2)$$

**AFM measurements:** Tapping mode AFM measurements were performed using a Bruker Veeco MultiMode IV at ambient conditions. The cantilever had a resonant frequency of 300 kHz and a force constant of 40 N/m. The scan rate was 1.0 Hz. The images were flattened/analyzed using Bruker NanoScope Analysis 1.70.

**Simulation methods:**

**Coarse-grained representation of DSA-Ph molecule:** The coarse-grained model for a DSA-Ph molecule (below, in Figure 4A) in this study consists of seven spherical beads, each representing a benzene ring. Type 1 beads represent the four peripheral benzene rings of DSA-Ph, type 2 beads represent the benzene rings connecting the central benzene ring to the peripheral rings and the type 3 bead represents the central benzene ring of DSA-Ph. Each coarse-grained bead interacts through a Lennard-Jones (LJ) potential with $\sigma_{bb}$=1.0 and $\varepsilon_{bb}$=1.0. A cutoff radius of 2.5 σ with a smooth decay starting at 2.4 σ is employed. To maintain the intramolecular structure, the beads within a molecule are connected by eight stiff bonds with harmonic stretching potentials ($L_b$=1.0, $k_b$=1000). The 1-2-3 bond angle is maintained at 150° with a harmonic bond-bending potential of spring constant $k_a$=1000. The 2-3-2 bond angle is maintained at 180 ° using the same spring constant. The non-bonded interaction potential is turned off for 1-2 (same as 2-3) and 1-3 bonded beads. No restriction was applied for the relative rotation along the longitudinal axis. The mass of each bead, $m_b$, is 1.0 in dimensionless units. The coarse-grained DSA-Ph molecule interacts with the substrate beads via a LJ potential. The reader is referred to the SI for detailed LJ parameters for anchoring and non-anchoring interfacial potentials.

**Coarse-grained representation of substrate:** The substrate is represented by spherical beads of mass $m_s$=1.0. The LJ interaction parameters between substrate beads are $\sigma_{ss}$=0.6 and $\varepsilon_{ss}$=0.1. To generate substrates with controlled roughness, the substrate surfaces are approximated by five superimposed two-dimensional Fourier functions, according to the equation:

$$z = \sum_{i=1}^{n=5} c_i cos(2\pi(u_i x + v_i y)) \qquad (3)$$

For each Fourier function, a two-dimensional random vector ($u_i$,$v_i$) is generated in the interval between zero and a specified wavelength unique to the type of surface being modeled. The scalar amplitude $c_i$ is randomly generated from an interval such that the average height of the final superimposed surfaces matches a specified value. To mimic the surface feature of $Si/SiO_2$ substrate[28], the Fourier wavelength is chosen to be 2 σ and the amplitude is chosen to be 0.7 σ. To initiate a substrate, 1500 substrate beads are generated with random locations in the $\hat{x}$-$\hat{y}$ plane. For each bead, using its $\hat{x}$-$\hat{y}$ coordinates, the $\hat{z}$ coordinate is calculated according to the superimposed Fourier function. Then a random number in the range from 0 to 0.1 σ is added to its $\hat{z}$ coordinate. The substrates are then minimized via the FIRE algorithm[29], with maximum displacement each step set to 0.01 σ. These substrate beads are then fixed in place for the entirety of the simulation using harmonic springs ($k_{sping}$=1000).

**Simulation box:** The simulation box size is 28 σ × 28 σ in the plane of the substrate ($\hat{x}$-$\hat{y}$) and 130 σ along the substrate normal ($\hat{z}$). Using the Van der Waal radius of a benzene ring[30] as an approximation for 1 σ, the box dimension is approximately 11 nm × 11 nm × 52 nm. Substrates are placed at both the top and the bottom of the simulation box. To enforce the solidity of the substrate surface, a continuous, repulsive potential is applied underneath the surface; this repulsive potential is implemented using a 12-6 LJ wall. The location and parameters of the LJ wall are chosen such that it does not contribute to intermolecular interactions at the surface. The box is periodic in the $\hat{x}$-$\hat{y}$ plane. A snapshot of the simulation box is shown in Figure S7.

**Deposition algorithm:** The simulation is performed using the Large-Scale Atomic/Molecular Massively Parallel Simulator (LAMMPS) package[31]. The simulated vapor deposition process is based on that reported earlier[15,32]. Each cycle consists of (i) the introduction of DSA-Ph molecules and diffusion to the top and or bottom substrates and (ii) cooling and diffusion along substrates. For (i), eight molecules are introduced to the simulation box consecutively, with two consecutive appearances being $5\times10^4$ steps apart. Upon appearance, the molecule is assigned a random position according to a spherical Gaussian distribution centered at the box geometric center. The velocity of the molecule is randomly drawn from a Maxwell-Boltzmann distribution at high temperature (T=1.0). Each bead within the molecule is initiated with the same velocity as the molecule. During (i), the newly introduced molecules have sufficient time to diffuse to either the top or bottom substrate surface. The molecules diffuse under a Langevin thermostat with a weak damping parameter ($t_{damp}$=5000 timesteps). During (ii), the newly deposited molecules are allowed to equilibrate on the substrates and are cooled to the substrate temperatures for $1\times10^6$ steps. The previously deposited molecules and the new molecules are all integrated under NVE ensemble. Throughout (i) and (ii), the substrates are maintained at a specified substrate temperature under Langevin thermostats with a standard damping parameter ($t_{damp}$=100 timesteps). The trajectory during the last $3\times10^5$ steps of (ii) is outputted and used for analysis. The integration timestep is 0.001 tau. Each simulation run generates two independent samples of deposited films (films on the top and bottom substrates of the simulation box).

**Analysis for z-dependent properties:** To obtain density and order parameter profiles along the substrate normal ($\hat{z}$), each simulation has 600 cycles, such that films with thicknesses of ~25 σ (~10nm) are grown on both top and bottom substrates. To calculate properties as a function of distance from the substrate, a deposited film is sliced into discrete 0.5 σ-thick slabs along $\hat{z}$. Molecules are assigned to slabs according to the $\hat{z}$ components of their centers of mass. For each slab, the number density is obtained by counting the number of beads in molecules that belongs to the slab. The orientational order parameter $P_2$ is defined as $P_2 = \langle \frac{3}{2}\cos^2\theta - \frac{1}{2} \rangle$, where $\theta$ is the angle between the longitudal axis of a molecule and the substrate normal, and $\langle \ldots \rangle$ denotes the average over time and over all molecules within the slab. For some comparisons, the bulk properties of the film are obtained by averaging over the middle region, where the distance from the substrate is between 10 σ and 15 σ. All $\hat{z}$-dependent profiles are averaged over all six independent films from three simulation runs.

**Results:**

GIWAXS patterns provide direct evidence that vapor-deposited DSA-Ph glasses as thin as 20 nm and as thick as 600 nm exhibit qualitatively the same average molecular packing. Shown in **Figure 1** are GIWAXS patterns from two DSA-Ph glasses of different thicknesses both deposited at 290K on a silicon substrate. $Q_z$ is the out of plane scattering vector and $Q_{xy}$ is the scattering vector in the plane. The colors represent scattered intensity (red = high scattered intensity, blue = low scattered intensity). For both diffraction patterns, at Q ~ 1.4 Å$^{-1}$, there is higher scattered intensity in the out-of-plane direction as compared to in-plane. The higher scattered intensity along $Q_z$ arises from a tendency towards face-on packing. Vapor-deposited DSA-Ph glasses varying in thickness, by a factor of 30, therefore exhibit qualitatively similar packing when deposited at the same substrate temperature (and deposition rate). The diffraction patterns in **Figure 1** also exhibit approximately the same peak position and width along Q (Figure S6), which is another indication of the similarity in packing in these glasses of different thickness.

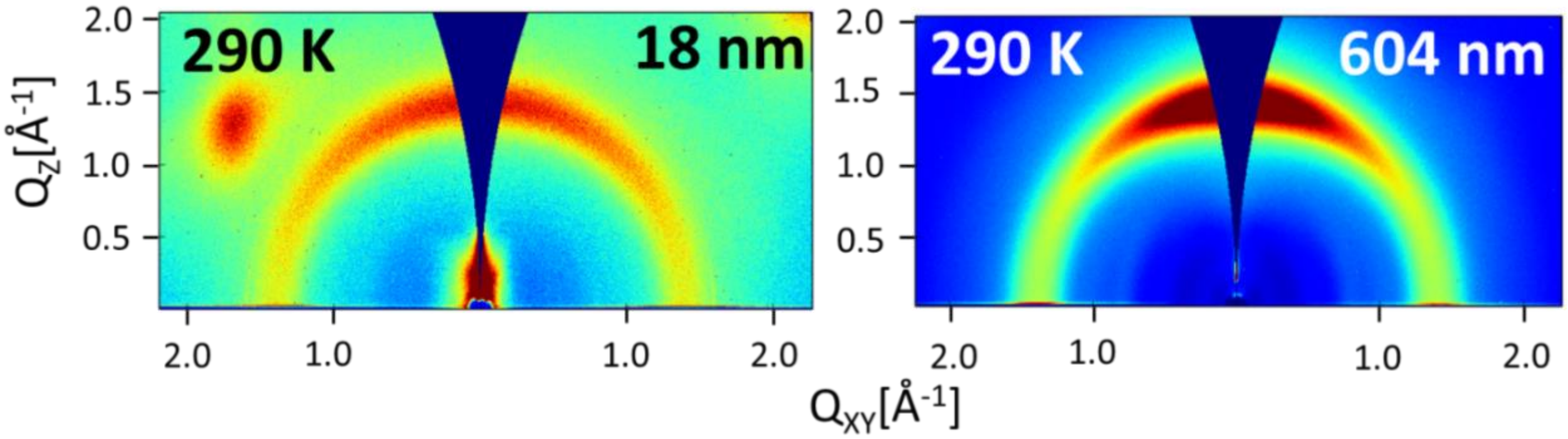

**Figure 1**: GIWAXS patterns from DSA-Ph films of thickness 18 nm and 604 nm both deposited at 290 K. Both the glasses scatter more strongly out of the plane (along $Q_z$) than in the plane (along $Q_{xy}$) at Q ~ 1.4 Å$^{-1}$; this arises from a tendency for face-on packing providing direct evidence that the packing is qualitatively similar in these glasses of different thickness. The feature in the left image at $Q_{xy}$ ~1.7 Å$^{-1}$ and $Q_z$ ~ 1.3 Å$^{-1}$ is diffuse scattering from the silicon substrate. Both the samples were deposited on a Si/$SiO_2$(2nm) substrate. The patterns were collected at an incidence angle of 0.14°, which is above the critical angle, and therefore representative of the bulk structure of the glass.

**Structure of ultrathin films as a function of deposition temperature**: 25 nm thick vapor-deposited glasses of DSA-Ph exhibit quantitatively similar structure as thicker films at all studied substrate temperatures. Shown in **Figure 2** is the Hermans order parameter ($S_{GIWAXS}$) as a function of substrate temperature for glasses of three different thicknesses. The $S_{GIWAXS}$ order parameter quantifies the scattering anisotropy at ~ 1.4 Å$^{-1}$. If all the scattered intensity was localized along $Q_z$ then $S_{GIWAXS}$=1; this occurs when there is perfect face-on packing. If all the scattered intensity was localized along $Q_{xy}$, $S_{GIWAXS}$=-0.5; this occurs when there is perfect edge-on or end-on packing. An $S_{GIWAXS}$=0 is consistent with isotropic packing.

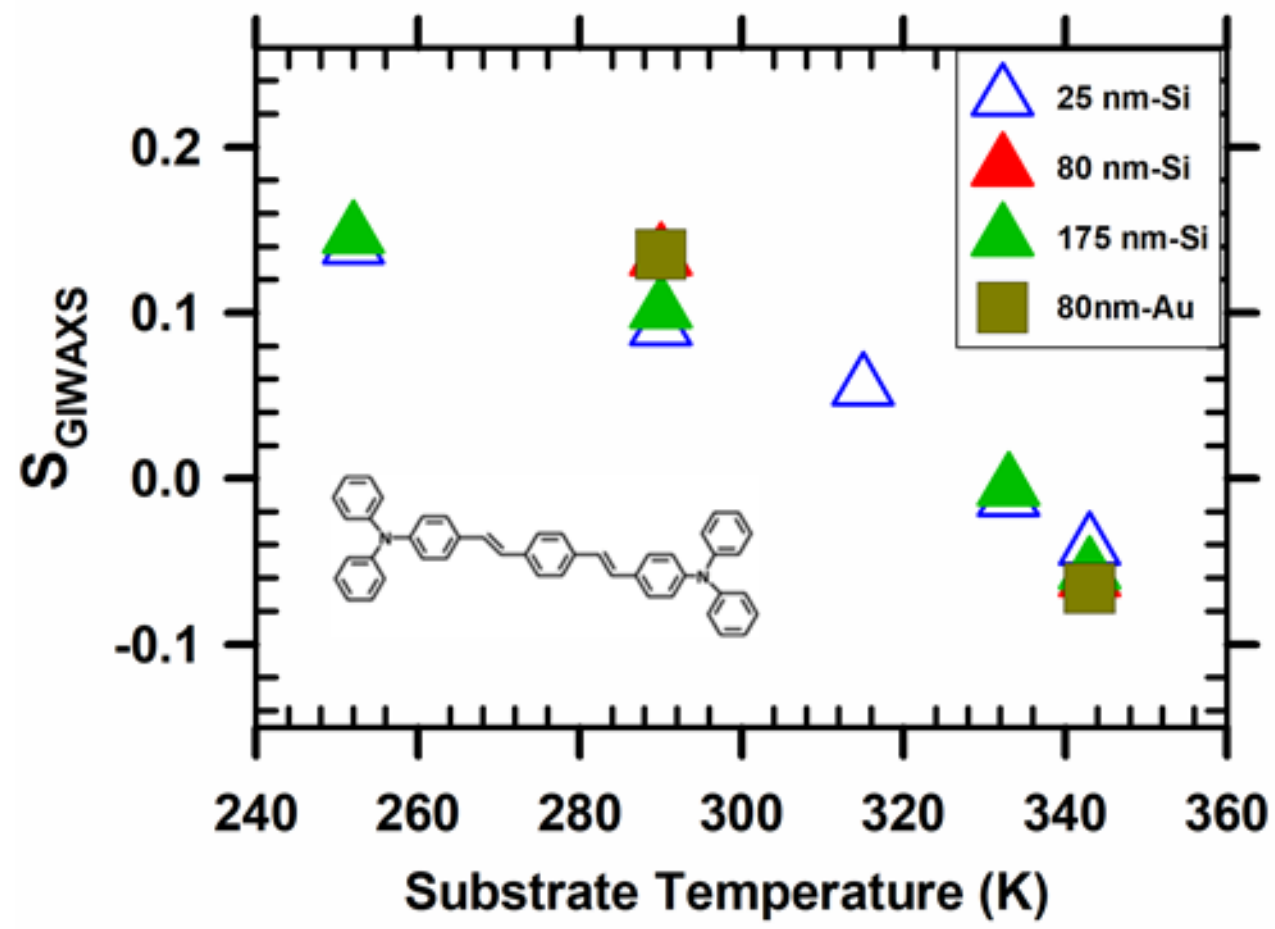

**Figure 2**: The GIWAXS order parameter as a function of substrate temperature for DSA-Ph glasses of three different thicknesses. 25 nm films exhibit similar structure as thicker films at all investigated substrate temperatures. The dark yellow squares represent order parameters for films deposited on gold. All other symbols represent depositions on Si (with ~ 2 nm of native oxide). The order parameters are evaluated from data obtained at an incidence angle of 0.14°. The molecular structure of DSA-Ph is shown in the inset

**Figure 2** shows that films of thickness ~25 nm (20-30 nm range), ~80 nm (70-90 nm), and ~ 175 nm (150-200nm) exhibit the same trend in the order parameter as previously observed for thicker films.[26] Films deposited at 252 K and 290 K exhibit a strong tendency towards face-on packing while films deposited at 343 K exhibit end-on packing. Quantitatively, **Figure 2** shows that 25 nm films of DSA-Ph exhibit order parameters nearly identical to thicker films at all studied substrate temperatures (raw diffraction patterns used to calculate order parameter for 25 nm and 175 nm films are presented in Figure S4). We find that the structure of DSA-Ph glasses deposited on gold and silicon substrates is almost identical, with this comparison including substrate temperatures that produce both face-on packing and end-on packing. Moreover, the two substrates utilized here have very different surface roughness (Fig S2). Previous work has suggested that substrate surface roughness influences the average structure of vapor-deposited glasses as thick as 100 nm[24]; this hypothesis clearly does not explain the data in **Figure 2**.

**Experimental estimate of interfacial length scale**: To understand over what distance the substrate influences the structure of vapor-deposited glasses of DSA-Ph, we deposited films of thickness 10-600 nm using a substrate temperature of 290 K. We focus on this particular substrate temperature since continuous films can be formed down to 10 nm thickness (Figure S1). In addition, 290 K is approximately room temperature and most OLED devices are fabricated with substrate temperature close to room temperature.

Shown in **Figure 3** is the Hermans order parameter, $S_{GIWAXS}$, as a function of film thickness at a deposition temperature of 290 K. We observe that the order parameter for all the films shown is positive, which indicates that films of thickness 10-600 nm exhibit a tendency towards face-on packing. We observe that films thinner than 20 nm exhibit a lower tendency towards face-on packing, with the thinnest film exhibiting roughly half the order parameter of the thickest films.

We observe that the order parameters for films deposited on gold are within error of those observed for deposition on silicon. Based on previous studies of conjugated molecules deposited on inorganic substrates, we expect the interactions of a conjugated molecule with gold and native oxide surfaces to be quite different[33]. The similarity between the structures of DSA-Ph films deposited on gold and silicon is an indication that preferred packing at the buried interface propagates over length scales far shorter than 25 nm. To estimate how much material near the substrate can have a distinct structure we fit the data in **Figure 3** to a two-layer model. We assume that there is an interfacial region that has isotropic packing, and that the rest of the film (the bulk) exhibits face-on packing. Mathematically, this model can be expressed as:

$$S_{GIWAXS}\,[h] = S_{bulk}\left[1 - \frac{\delta}{h}\right] \qquad (4)$$

Here $S_{GIWAXS}$ is the observed order parameter, $S_{bulk}$ is the order parameter of the bulk region, δ is the thickness of the isotropic interfacial region (with $S_{GIWAXS}$=0) and h is the total thickness of the film. $S_{bulk}$ and δ are fit parameters in the model.

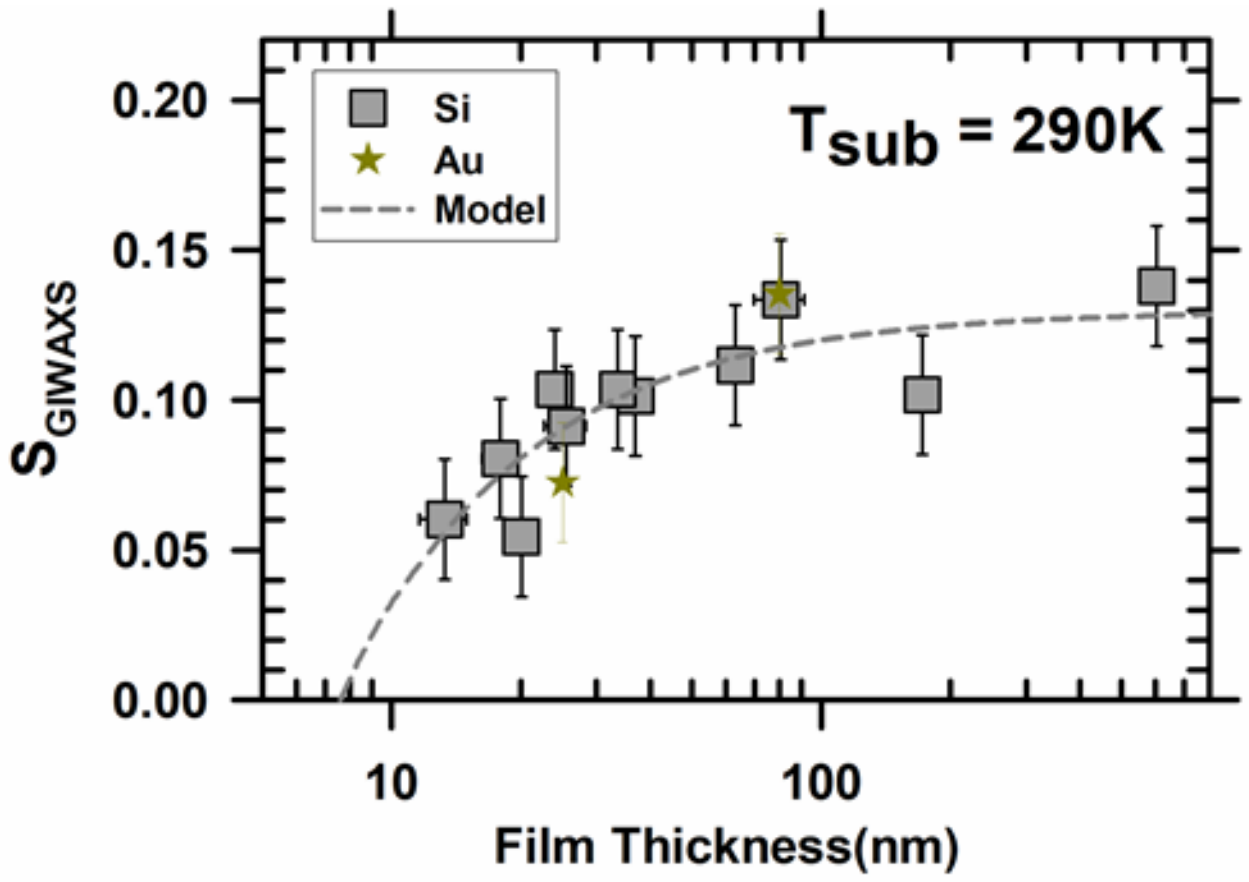

**Figure 3**: The GIWAXS order parameter plotted as a function of film thickness for films deposited at $T_{sub}$=290 K. The gray squares represent films deposited on $Si/SiO_2$ (2 nm) and the gold stars

represent films deposited on gold. The dashed line is a fit to a two-layer model for films deposited on $Si/SiO_2$. Based on the two-layer fit we estimate the size of the interfacial region with different packing to be 7.6 ± 1.6 nm.

Our two-layer model provides an estimate (see below) for how much material near the substrate can exhibit qualitatively different packing. A similar approach to that taken in **Figure 3** has been used to extract thickness of interfacial layers for partially crystalline films of semiconducting polymers[34]. At a deposition temperature of 290 K, our model provides an estimate of 7.6 ± 1.6 nm for the thickness of the interfacial region with perturbed packing. We have insufficient data points for films deposited on gold to make such an estimate, but the similarity in structure for films deposited on the two substrates is reasonably interpreted as a similar length scale for films deposited on gold. In Figure S3, we show a similar trend with film thickness for films deposited at $T_{sub}$=252 K, with an interfacial length scale of 4.7 ± 1.6 nm obtained from the two-layer model. Depositions on different substrates and at different deposition temperatures all support the conclusion of a nanometer-scale influence of the substrate on the structure of vapor-deposited glasses.

We note that the ~ 8 nm interfacial length-scale extracted in **Figure 3** at $T_{sub}$=290 K, follows from several simplifying assumptions. For simplicity, we assume in our model that the buried interface is isotropic. Edge-on or end-on packing at the buried interface would also explain the lowering of the order parameter and, in those cases, the length scale would be less than 8 nm. Moreover, based on data collected below the critical angle (Figure S5) we know that the free surface also contributes to the observed lowering of order in thinner films (at $T_{sub}$=290 K). With our experimental data, it is not possible to uniquely separate or distinguish the contributions of the free surface and the buried interface to the lowering of order in the thinnest film. To overcome the ambiguities associated with our experimental estimate for how far the substrate can perturb the structure of a vapor-deposited glass, we turn to molecular dynamics simulations.

**MD simulations of structure near solid interface**: To gain further insight into the interfacial structure of PVD glasses we perform molecular dynamics simulations on vapor-deposited glasses of a coarse-grained model of DSA-Ph. In previous work, coarse-grained Lennard-Jones models have been successful in reproducing experimental results for anisotropy in thick PVD glasses and in explaining the basic physics of structure formation in vapor-deposited glasses[15]. Molecular dynamics simulations provide the advantage that the substrate interface can be directly probed at a sub-nanometer length scale. Moreover, molecule-substrate interactions in simulations can be tuned in a precise and straightforward manner. A simulation provides the opportunity to create a specific orientation at the buried interface by choosing the appropriate potential; how far the induced order propagates into the glass provides a direct estimate of the length-scale over which the substrate perturbs the structure of a vapor-deposited molecular glass.

**Figure 4** depicts the coarse-grained representation of DSA-Ph used in our simulations and shows how DSA-Ph molecules interact with substrates that have different anchoring potentials. The molecular model is chosen to capture the molecular geometry of DSA-Ph. DSA-Ph has seven aromatic rings; we therefore construct a representation of DSA-Ph with seven Lennard-Jones beads. The beads are numbered based on their position in the molecule. In **Figure 4B**, the orientation of molecules with two different molecule-substrate interaction potentials are shown at a substrate temperature of 0.68. For the "non-anchoring" potential, the Lennard-Jones parameters describing interactions with the substrate are the same for all the beads of the molecule. The snapshot with the non-anchoring potential shows molecules are horizontally-oriented; such an orientation helps molecules best minimize their energy at the free surface. For the "vertically anchoring" potential the interaction between the terminal beads of the molecule (see Supplemental information Table 1) and the substrate beads are more favorable than other interactions, causing the molecule to adopt a vertical orientation at the buried interface. The parameters for molecule-substrate interactions are specified in SI Table 1.

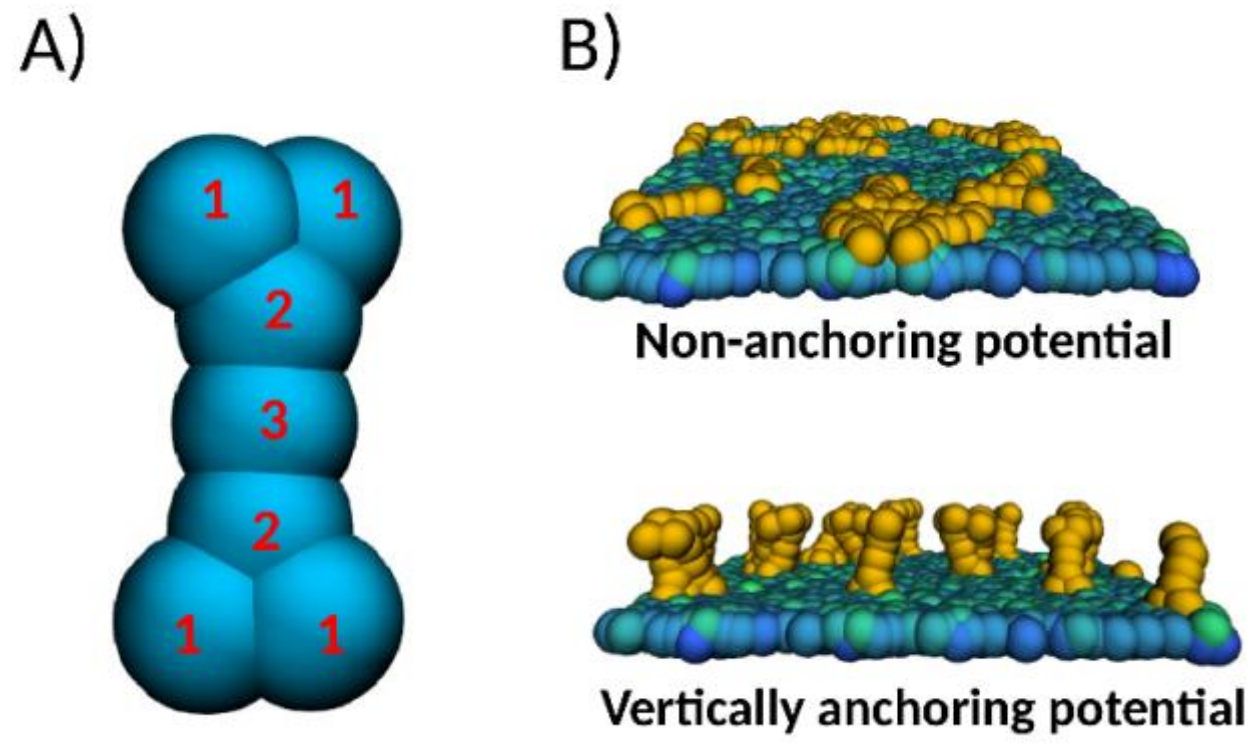

**Figure 4**: Schematic of a coarse-grained DSA-Ph molecule (A). Snapshots of coarse-grained molecules on substrates with two different interfacial potentials(B). In (B), substrate atoms are colored based on their position in z-direction (substrate normal).

Shown in **Figure 5A** and **B** are the orientational order parameter, $P_2$, and density as a function of depth into the film for coarse-grained DSA-Ph glasses deposited at 0.68 with the two different substrate-molecule interaction potentials. The substrate interface is located at the left axis of these graphs. $P_2$ is a measure of molecular orientation (defined in the methods section). A $P_2$ order parameter of 1 would mean there is perfect vertical orientation of the long axes of the molecules and a value of -0.5 would mean that all the long axes are perfectly horizontally oriented. The two plots (**5A** and **B**) show that beyond a distance of 5 σ the glass structure is independent of interactions at the buried interface. By comparing the lengths of the actual and coarse-grained molecule, a length of 5 σ corresponds to approximately 2.8 nm. We can quantitatively compare the structure in the bulk region of the film, defined as the material between 10-15 σ from the substrate. Under the non-anchoring interfacial potential, the order parameter is -0.306±0.014 and density is 0.926±0.033 in the bulk. Under the vertically anchoring interfacial potential, the order parameter is -0.274±0.030 and density is 0.932±0.050 in the bulk.

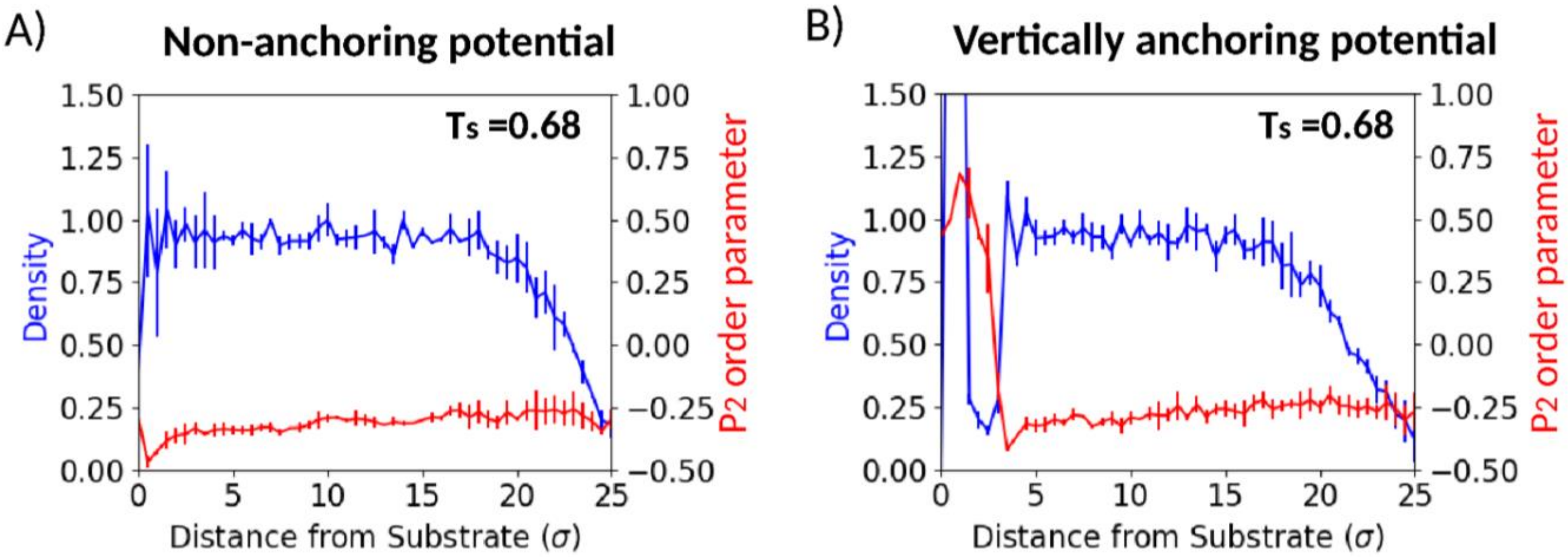

**Figure 5**: Density (beads/$\sigma^3$) and $P_2$ orientation order parameter as a function of distance from substrate, for simulations in which DSA-Ph is deposited onto substrates with A) non-anchoring potential and B)vertically anchoring potential . The orientation order parameter is plotted as a red solid line and density is plotted as a blue solid line. The two panels show that beyond 5 σ (~ 3 nm) the structure of the PVD glass is independent of interactions at the buried interface. For these simulations, the substrate temperature was 0.68. Error bars represent one standard deviation from six independent replicas.

The ~ 3 nm interfacial length scale obtained from simulations is consistent with the upper bound of ~8 nm extracted from x-ray scattering measurements. We note that these length scales were obtained by considering different but related observables. The GIWAXS order parameter utilized in the experiments characterizes the direction in reciprocal space where there is excess scattered intensity while the $P_2$ order parameter characterizing the simulations describes the average orientation of the long axes of the molecules. As a result, while perfect face-on packing would produce a GIWAXS order parameter of 1, it would correspond to a $P_2$ value of -0.5. We chose to calculate the $P_2$ order parameter in simulations because it can be precisely measured as a function of depth in the film. As previous studies of thick films have established that there is a strong one-to one correlation between the $P_2$ order parameter for the long axis and the GIWAXS order parameter,[35] it is reasonable to compare the length-scales extracted from these two quantities.

In **Figure 6,** the orientation order parameter is plotted as a function of depth into the film, for a simulation where the substrate-molecule interaction produces highly vertical molecular orientation at the buried interface. It is clear from **Figure 6** that after 5 σ (~3 nm) the molecular orientation is determined by the substrate temperature during deposition and not substrate-molecule interactions. The ~3 nm length scale over which the substrate perturbs the structure of a vapor-deposited computer glass is in qualitative agreement with the nanometer-scale interfacial length inferred from **Figure 2 & 3**; both simulations and experiments establish that the substrate perturbs the structure of a vapor-deposited glass for less than 8 nm. (In **Figure 6**, the profile at T=0.68 behaves differently from profiles at other substrate temperatures near the free surface. At this low substrate temperature, the surface mobility is very limited; long simulations of the deposited film allow the surface to equilibrate and under these conditions the surface orientation evolves towards that shown for the other substrate temperatures.)

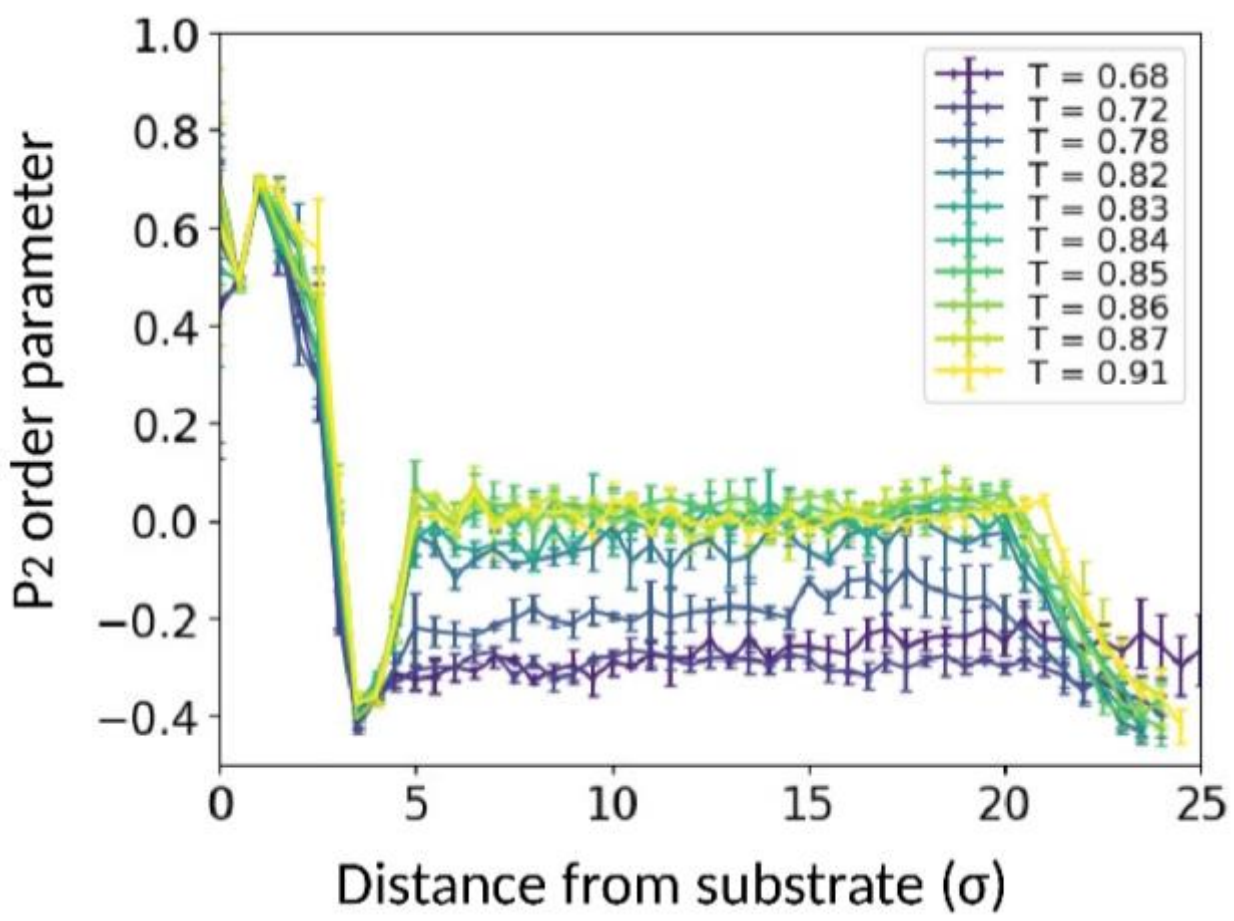

**Figure 6**: Simulation results showing the $P_2$ orientation order parameter for DSA-Ph molecules as a function of distance from substrate at various substrate temperatures. These results were obtained with the vertically anchoring potential and the results are averaged over six independent replicas. Error bars represent one standard deviation.

While simulation and experiment agree that a solid substrate influences the structure of a vapor-deposited glass for less than 8 nm, the quantitative difference between the length scale estimates (3 nm vs 8 nm) requires further discussion. The experimental estimate of 8 nm, within the context of the two-layer model, is expected to be an upper-bound; as mentioned above, the 8 nm estimate includes contributions from structural differences at the free surface. The deposition rate used in simulations is seven orders of magnitude greater than that used experimentally. The slower deposition rate in experiments allows more time for dewetting and this provides a further possible explanation for the larger length scale extracted from the experimental data. In simulations, a continuous film is formed after the first layer is deposited; in experiments it is likely that several molecular layers are required for a continuous film to form[36]. For depositions at 290 K, we know that 10 nm films are continuous and smooth; however, this does not rule out the possibility that dewetting occurs during the first ~ 5 nm of the deposition. As the material rearranged by dewetting is likely isotropic, this effect would give rise to a larger interfacial length scale. Despite these differences in experimental and simulation deposition conditions, both approaches observe an effect of the solid substrate extending less than 10 nm. This provides strong evidence that this conclusion is robust under deposition conditions where a continuous film is formed by 10 nm. As discussed below, this requirement for continuous thin films is consistent with typical deposition conditions, for example, those utilized for OLEDs.

## Discussion:

**Comparison to previous work:** Our experiments go beyond previous studies that have looked at the influence of solid substrates on PVD glass structure[24,25]. To the best of our knowledge, this is the first study which quantifies the structure of a vapor-deposited glass as thin as ~13 nm. Moreover, while previous studies[25] have varied film thickness by a factor of ~5, in our study we have varied thickness by a factor of ~ 45 (**Figure 3**). By studying glass structure across this much broader range of film thickness, we can confidently estimate the length scale of interfacial perturbations. In addition to film thickness, we vary substrate temperature and the substrate identity. Across this broad range of conditions, we find that the perturbation of the solid substrate on vapor-deposited glass structure is a few nanometers. Moreover, computer simulations of the deposition process presented here support this conclusion. These simulations are the first to vary the anchoring of deposited molecules at the substrate while holding all other variables fixed. Over a wide range of substrate temperature, the simulation results indicate that the solid substrate influences PVD glass structure for a few nanometers.

In the literature, results from investigations of the effect of the substrate on PVD glass structure fall into two broad categories. One group of results indicates that the substrate influences the average structure of glasses as thick as 100 nm while another group of results estimates the substrate influences glass structure over a length scale at least an order of magnitude smaller. Yokoyama et al found that 100 nm films of vapor-deposited glasses deposited on sapphire, fused silica, and ITO glass exhibit different average structures, implying that the substrate can have an effect on glass structure over long length scales[24]. Along similar lines Yoshizaki et al[37] found that 200 nm thick Alq3 films deposited on aluminum exhibit roughly half the polar order compared to films grown on gold substrates. On the other hand, Sakai et al[25] showed that films of thickness 20-100 nm exhibit the same average structure; moreover, this study found no difference when films were deposited on silicon or fused silica. Two previous simulation efforts are consistent with the view that that substrate influence on glass structure is limited to less than ~ 5 nm[38,39].

Our experiments allow us to test the hypothesis proposed by Yokoyama et al[24] that the substrate roughness can influence the structure of vapor-deposited glasses over long distances. Yokoyama et al[24] report that the substrate influences the average structure of films as thick as 100 nm. Based on optical absorbance, these authors conclude that the average molecular orientation of BSB-Cz in 100 nm films deposited on sapphire is different than the orientation in 100 nm films deposited on fused silica. They identify substrate roughness as the cause for this difference; they report that the RMS roughness of their fused silica substrate is 0.9 nm, while the RMS roughness of sapphire is 0.2 nm[24]. This line of reasoning does not explain our results. The RMS roughness of our gold surface is ~1 nm, while the RMS roughness of $Si/SiO_2$ is ~ 0.2 nm (AFM patterns are shown in Fig S2). While the difference in roughness between gold and silicon is very similar to the difference in roughness between the aforementioned substrates used by Yokoyama et al., we find that DSA-Ph films deposited on gold and silicon exhibit the same structure within experimental error (**Figure 2 & 3**), even for films as thin as 25 nm.

With regard to the observations of Yoshizaki et al[37], a number of studies on polar ordering in vapor-deposited Alq3 glasses support the view that a long-range perturbation of the substrate is unlikely to be a general result. Alq3 glasses deposited on five different solid substrates[37,40,41,42,43] have the same surface potential for the same film thickness, consistent with the idea that the substrate does not have a long-range effect on glass structure. A sixth substrate, aluminum, showed a significantly different surface potential.[37] While we do not know why an aluminum substrate would affect polar ordering in Alq3, it is possible that this result was influenced by crystallization or de-wetting during deposition.

The short ranged-influence of the substrate reported here for vapor-deposited glasses of organic semiconductors stands in contrast to the behavior of several other important materials systems. For inorganic semiconductors, the substrate is used to drive epitaxial growth over the entire thickness of the film. For crystalline organic semiconductors such as pentacene, the substrate induced phase can propagate for ~100 nm[23]. In polymers, the substrate has been shown to have a ~100 nm influence on dynamics[44]. In liquid-crystals, alignment layers can induce preferential molecular orientation over hundreds of nanometers[45]; this difference in the influence of the substrate occurs in spite of the similarity in structure of vapor-deposited glasses and liquid-crystalline phases[16]. Establishing this short length-scale helps identify a key fundamental difference between the order in vapor-deposited glasses and other types of technologically important materials. Moreover, a quantitative estimate for this length-scale is important for applications which utilize vapor-deposited glasses, such as OLEDs.

The interfacial length scale of a few nanometers observed in our study can be rationalized based on our understanding of molecular glasses. The short-ranged nature of van der Waals interactions in combination with deposition into an amorphous state, makes it likely that the substrate perturbs glass structure over only nanometer length scales. To illustrate the unique features of amorphous deposition, we first remind the reader how crystals propagate order over large distances. At the surface of a growing crystal, there is a deep free energy minimum guiding the attachment of new molecules; molecular attachment that does not propagate the structure of the existing crystal is unlikely due to the high energy cost of defects (such as dislocations and grain boundaries). In contrast, because there are many local packing arrangements with similar energies in glasses, there are no easily identified defects. At the surface of a growing glass, each new set of molecules has many packing arrangements of similar energy, and thus only a few layers of growth are required for the glass to lose memory of the underlying structure. The simulations presented here reinforce this physical picture. By introducing strong vertical anchoring at a substrate temperature known to produce horizontal molecular orientation, we monitor in our simulations the competition between substrate-induced ordering and the driving force at the free surface to produce horizontal molecular orientation[15]. At all substrate temperatures, these simulations show that molecular order induced by the solid substrate propagates only a few nanometers into the deposited glass.

Our experiments and simulations indicate that vapor-deposited glasses cannot propagate order more than a few nanometers from a substrate unless additional mechanisms involving larger length scales are important. One example of such an additional mechanism would be dewetting during deposition, which could influence the film structure up to the length scale required to deposit a continuous film[36,46]. To avoid this effect, in this work we have restricted our attention to deposition conditions that produce smooth, continuous films as thin as 10 nm. Continuous films are required for OLED devices making the deposition conditions we focus on technologically relevant. Another scenario not considered here is deposition slightly below $T_g$, where if the deposition rates are slow enough the entire bulk of the film can equilibrate during deposition. We focus in our study on substrate temperature well below the glass transition temperature of DSA-Ph, where mobility is restricted to the top surface layers[15], and consequently only the surface can equilibrate. As most OLED devices are fabricated at room temperature and as most organic semiconductors have a $T_g$ at least 30 K above room temperature, the deposition temperatures considered in this study span the typical range of $T_{sub}/T_g$ used for device fabrication.

The conclusion of a nanometer-scale interfacial length for PVD glasses can be directly tested by other experimental techniques, and such measurements can provide further understanding of structure at buried interfaces. Polarized soft-x-ray reflectivity can be used to depth profile molecular orientation in organic thin films[47] ; in addition to measuring the length scale over which the substrate influences structure, soft x-ray reflectivity can be used to characterize the structure at the buried interface. NEXAFS spectroscopy can also be utilized to this end. Performing NEXAFS in surface-sensitive electron-yield modes on delaminated samples might allow the structure of the buried interface[48] to be probed. While our experiments observe less average order in the thinnest films, whether this is due to isotropic, edge-on or end-on packing at the substrate is unknown. We expect that the utilization of soft-x-ray reflectivity and NEXAFS spectroscopy can fill this void in our understanding. Measuring UPS and XPS spectra on PVD glasses of DSA-Ph as a function of film thickness could elucidate how the structural perturbations near the substrate observed here influence electronic structure. Previous studies have measured UPS and XPS spectra as a function of thickness for other molecules[49,50]; similar measurements on DSA-Ph would allow a direct correlation of electronic structure(eg..HOMO level, energetic disorder) with molecular packing.

**Implications for organic electronics:** The structure at the interface of a vapor-deposited organic semiconductor glass and an inorganic substrate has not been directly examined, despite its importance for organic electronic applications. The molecular packing at the interface of an organic semiconductor and inorganic electrode will influence how efficiently charge is injected into the material in OLED devices[11]. Previous device studies suggest that changing the deposition temperature of a 5 nm PVD glassy layer at the interface with indium tin oxide can change the required driving voltage by up to a factor of two[51]; this difference in devices where the 5 nm layer near the substrate is prepared at different temperatures is likely due the differences in interfacial molecular packing. Although to a lesser extent compared to OLEDs, vapor-deposited glasses have also been utilized to fabricate OFET devices[52]; here performance will depend critically on the structure at the interface of the organic-semiconductor and the dielectric[53]. As a first step towards understanding this interface, we establish here the length scale over which the substrate can influence PVD glass structure. Although the structure at the interface will depend on the choice of the molecule and substrate, we expect the observation of a nanometer scale effect of the substrate on the structure to be transferable across different systems. The similarity of PVD glass structure observed for deposition onto different substrates, observed both experimentally and computationally, provide support for this. Our results are also relevant for understanding why the elastic moduli of PVD glasses depends upon film thickness below 20 nm[54,55]; the moduli of thin films could be significantly influenced by the glass structure near the buried interface.

Finally, our results also demonstrate that, beyond 3 to 8 nm near the solid substrate, the unique anisotropic glassy packing motifs[14] that are accessible through PVD can be produced on arbitrary substrates.

**Conclusions**

The length scale over which the substrate perturbs the structure of PVD glasses of DSA-Ph is found to be 3 to 8 nm. The altered structure near the substrate is significant for OLED devices which often use layers as thin as 10 nm. The interfacial length scale determined in this work will enable future direct investigation of the structure of vapor-deposited organic semiconductors near solid substrates.

AUTHOR INFORMATION

**Corresponding Author**

kbagchi@wisc.edu

**Author Contributions**

The manuscript was written through contributions of all authors. All authors have given approval to the final version of the manuscript.

ASSOCIATED CONTENT

**Supplemental information:**

AFM image of 10 nm DSA-Ph film deposited on a Si/$SiO_2$(2 nm) substrate is shown in Figure S1. AFM images of bare Si/$SiO_2$(2 nm) and gold substrates are shown in Figure S2. $S_{GIWAXS}$ order parameter as a function of film thickness at $T_{sub}$=252 K is shown in Figure S3. Diffraction patterns of 25 nm and 175 nm films deposited at various deposition temperatures is shown in Figure S4.

The $S_{GIWAXS}$ order parameter at two incidence angles for various deposition temperatures is shown in Figure S5. The one-dimensional scattering profiles for the patterns in Figure 1 is shown in Figure S6. The simulation box is shown in Figure S7. Figure S8 shows determination of $T_g$ in simulations and Figure S9 shows characterization of interface molecular orientation.

**Funding Sources**

The experimental portion of this work was supported by the US Department of Energy, Office of Basic Energy Sciences, Division of Materials Sciences and Engineering, Award DE-SC0002161. The authors gratefully acknowledge use of facilities and instrumentation supported by NSF through the University of Wisconsin Materials Research Science and Engineering Center (DMR-1720415). Use of the Stanford Synchrotron Radiation Lightsource, SLAC National Accelerator Laboratory, is supported by the U.S. Department of Energy, Office of Science, Office of Basic Energy Sciences, under Contract DE-AC02-76SF00515. The molecular simulations presented in this work were carried out at Argonne National Laboratory with support from the Department of Energy, Office of Science, Basic Energy Sciences, Division of Materials Science and Engineering

**ACKNOWLEDGMENT**

We thank Christopher J. Tassone and Tim J. Dunn for assistance at BL 11-3. We thank Richard Noll for assistance with operating the Leica EM ACE 600 Coater.

## REFERENCES

(1) Williams, C. T.; Beattie, D. A. Probing Buried Interfaces with Non-Linear Optical Spectroscopy. *Surf. Sci.* **2002**, *500*, 545–576.

(2) Chen, Z. Understanding Surfaces and Buried Interfaces of Polymer Materials at the Molecular Level Using Sum Frequency Generation Vibrational Spectroscopy. *Polym. Int.* **2007**, *56*, 577–587.

(3) Jones, A. O. F.; Chattopadhyay, B.; Geerts, Y. H.; Resel, R. Substrate-Induced and Thin-Film Phases: Polymorphism of Organic Materials on Surfaces. *Adv. Funct. Mater.* **2016**, *26*, 2233–2255.

(4) Yu, C.-J.; Richter, A. G.; Datta, A.; Durbin, M. K.; Dutta, P. Molecular Layering in a Liquid on a Solid Substrate: An X-Ray Reflectivity Study. *Phys. B Condens. Matter* **2000**, *283*, 27–31.

(5) Ocko, B. M. Smectic-Layer Growth at Solid Interfaces. *Phys. Rev. Lett.* **1990**, *64*, 2160.

(6) Kauzmann, W. The Nature of the Glassy State and the Behavior of Liquids at Low Temperatures. *Chem. Rev.* **1948**, *43*, 219–256.

(7) Ediger, M. D.; Harrowell, P. Perspective: Supercooled Liquids and Glasses. *J. Chem. Phys.* **2012**, *137*, 80901.

(8) Jou, J.-H.; Kumar, S.; Agrawal, A.; Li, T.-H.; Sahoo, S. Approaches for Fabricating High Efficiency Organic Light Emitting Diodes. *J. Mater. Chem. C* **2015**, *3*, 2974–3002.

(9) Ràfols-ribé, J.; Will, P.; Hänisch, C.; Gonzalez-silveira, M.; Lenk, S.; Rodríguez-viejo, J.; Reineke, S. High-Performance Organic Light-Emitting Diodes Comprising Ultrastable Glass Layers. *Sci. Adv.* **2018**, *4*.

(10) Menke, S. M.; Luhman, W. A.; Holmes, R. J. Tailored Exciton Diffusion in Organic Photovoltaic Cells for Enhanced Power Conversion Efficiency. *Nat. Mater.* **2013**, *12*, 152–157.

(11) Chen, W.; Huang, H.; Chen, S.; Huang, Y. L.; Gao, X. Y.; Wee, A. T. S. Molecular Orientation-Dependent Ionization Potential of Organic Thin Films. *Chem. Mater.* **2008**, *20*, 7017–7021.

(12) Yokoyama, D. Molecular Orientation in Small-Molecule Organic Light-Emitting Diodes. *J. Mater. Chem.* **2011**, *21*, 19187−19202.

(13) Schmidt, T. D.; Lampe, T.; R, D. S. M.; Djurovich, P. I.; Thompson, M. E.; Brütting, W. Emitter Orientation as a Key Parameter in Organic Light-Emitting Diodes. *Phys. Rev. Appl.* **2017**, *8*, 037001.

(14) Ediger, M. D.; de Pablo, J. J.; Yu, L. Anisotropic Vapor-Deposited Glasses: Hybrid Organic Solids. *Acc. Chem. Res.* **2019**, *52*, 407–414.

(15) Dalal, S. S.; Walters, D. M.; Lyubimov, I.; de Pablo, J. J.; Ediger, M. D. Tunable Molecular Orientation and Elevated Thermal Stability of Vapor-Deposited Organic Semiconductors. *Proc. Natl. Acad. Sci.* **2015**, *112*, 4227–4232.

(16) Bishop, C.; Thelen, J. L.; Gann, E.; Toney, M. F.; Yu, L.; DeLongchamp, D. M.; Ediger, M. D. Vapor Deposition of a Nonmesogen Prepares Highly Structured Organic Glasses. *Proc. Natl. Acad. Sci.* **2019**, *116*, 21421–21426.

(17) Bagchi, K.; Jackson, N. E.; Gujral, A.; Huang, C.; Toney, M. F.; Yu, L.; de Pablo, J. J.; Ediger, M. D. Origin of Anisotropic Molecular Packing in Vapor-Deposited Alq3 Glasses. *J. Phys. Chem. Lett.* **2018**, *10*, 164–170.

(18) Bishop, C.; Gujral, A.; Toney, M. F.; Yu, L.; Ediger, M. D. Vapor-Deposited Glass Structure Determined by Deposition Rate-Substrate Temperature Superposition Principle. *J. Phys. Chem. Lett.* **2019**, *10*, 3536–3542.

(19) Mannsfeld, S. C. B.; Virkar, A.; Reese, C.; Toney, M. F.; Bao, Z. Precise Structure of Pentacene Monolayers on Amorphous Silicon Oxide and Relation to Charge Transport. *Adv. Mater.* **2009**, *21*, 2294–2298.

(20) Himmelberger, S.; Dacuña, J.; Rivnay, J.; Jimison, L. H.; McCarthy-Ward, T.; Heeney, M.;

McCulloch, I.; Toney, M. F.; Salleo, A. Effects of Confinement on Microstructure and Charge Transport in High Performance Semicrystalline Polymer Semiconductors. *Adv. Funct. Mater.* **2013**, *23*, 2091–2098.

(21) Xu, J.; Diao, Y.; Zhou, D.; Mao, Y.; Giri, G.; Chen, W. Probing the Interfacial Molecular Packing in TIPS-Pentacene Organic Semiconductors by Surface Enhanced Raman Scattering. **2014**, 2985–2991.

(22) Kline, R. J.; McGehee, M. D.; Toney, M. F. Highly Oriented Crystals at the Buried Interface in Polythiophene Thin-Film Transistors. *Nat. Mater.* **2006**, *5*, 222.

(23) Bouchoms, I. P. M.; Schoonveld, W. A.; Vrijmoeth, J.; Klapwijk, T. M. Morphology Identification of the Thin Film Phases of Vacuum Evaporated Pentacene on SiO2 Substrates. *Synth. Met.* **1999**, *104*, 175–178.

(24) Yokoyama, D.; Setoguchi, Y.; Sakaguchi, A.; Suzuki, M.; Adachi, C. Orientation Control of Linear-Shaped Molecules in Vacuum-Deposited Organic Amorphous Films and Its Effect on Carrier Mobilities. *Adv. Funct. Mater.* **2010**, *20*, 386–391.

(25) Sakai, Y.; Shibata, M.; Yokoyama, D. Simple Model-Free Estimation of Orientation Order Parameters of Vacuum-Deposited and Spin-Coated Amorphous Films Used in Organic Light-Emitting Diodes. *Appl. Phys. Express* **2015**, *8*, 096601.

(26) Bagchi, K.; Gujral, A.; Toney, M. F.; Ediger, M. D. Generic Packing Motifs in Vapor-Deposited Glasses of Organic Semiconductors. *Soft Matter* **2019**, *15*, 7590–7595.

(27) Baker, J. L.; Jimison, L. H.; Mannsfeld, S.; Volkman, S.; Yin, S.; Subramanian, V.; Salleo, A.; Alivisatos, A. P.; Toney, M. F. Quantification of Thin Film Crystallographic Orientation Using X-Ray Diffraction with an Area Detector. *Langmuir* **2010**, *26*, 9146–9151.

(28) Khalilov, U.; Neyts, E. C.; Pourtois, G.; Van Duin, A. C. T. Can We Control the Thickness of Ultrathin Silica Layers by Hyperthermal Silicon Oxidation at Room Temperature? *J. Phys. Chem. C* **2011**, *115*, 24839–24848.

(29) Bitzek, E.; Koskinen, P.; Gähler, F.; Moseler, M.; Gumbsch, P. Structural Relaxation Made

Simple. *Phys. Rev. Lett.* **2006**, *97*, 1–4.

(30) Von Lilienfeld, O. A.; Andrienko, D. Coarse-Grained Interaction Potentials for Polyaromatic Hydrocarbons. *J. Chem. Phys.* **2006**, *124*.

(31) Plimpton, S. Fast Parallel Algorithms for Short-Range Molecular Dynamics. *J. Comput. Phys.* **1995**, *117*, 1–19.

(32) Reid, D. R.; Lyubimov, I.; Ediger, M. D.; de Pablo, J. J. Age and Structure of a Model Vapor-Deposited Glass. *Nat. Commun.* **2016**, *7*, 1–9.

(33) Käfer, D.; Ruppel, L.; Witte, G. Growth of Pentacene on Clean and Modified Gold Surfaces. *Phys. Rev. B* **2007**, *75*, 85309.

(34) Jimison, L. H.; Himmelberger, S.; Duong, D. T.; Rivnay, J.; Toney, M. F.; Salleo, A. Vertical Confinement and Interface Effects on the Microstructure and Charge Transport of P3HT Thin Films. *J. Polym. Sci. Part B Polym. Phys.* **2013**, *51*, 611–620.

(35) Gujral, A.; O'Hara, K. A.; Toney, M. F.; Chabinyc, M. L.; Ediger, M. D. Structural Characterization of Vapor-Deposited Glasses of an Organic Hole Transport Material with X-Ray Scattering. *Chem. Mater.* **2015**, *27*, 3341–3348.

(36) Farahzadi, A.; Niyamakom, P.; Beigmohamadi, M.; Meyer, N.; Keiper, D.; Heuken, M.; Ghasemi, F.; Tabar, M. R. R.; Michely, T.; Wuttig, M. Stochastic Analysis on Temperature-Dependent Roughening of Amorphous Organic Films. *EPL (Europhysics Lett.* **2010**, *90*, 10008.

(37) Yoshizaki, K.; Manaka, T.; Iwamoto, M. Large Surface Potential of Alq3 Film and Its Decay. *J. Appl. Phys.* **2005**, *97*, 23703.

(38) Youn, Y.; Yoo, D.; Song, H.; Kang, Y.; Kim, K. Y.; Jeon, S. H.; Cho, Y.; Chae, K.; Han, S. All-Atom Simulation of Molecular Orientation in Vapor-Deposited Organic Light-Emitting Diodes. *J. Mater. Chem. C* **2018**, *6*, 1015–1022.

(39) Lyubimov, I.; Antony, L.; Walters, D. M.; Rodney, D.; Ediger, M. D.; Pablo, J. J. De.

Orientational Anisotropy in Simulated Vapor-Deposited Molecular Glasses. *J. Chem. Phys.* **2015**, *143*, 094502.

(40) Okabayashi, Y.; Ito, E.; Isoshima, T.; Hara, M. Positive Giant Surface Potential of Tris(8-Hydroxyquinolinolato)Aluminum (Alq3) Film Evaporated onto Backside of Alq3 Film Showing Negative Giant Surface Potential. *Appl. Phys. Express* **2012**, *5*, 055601.

(41) Isoshima, T.; Okabayashi, Y.; Ito, E.; Hara, M.; Chin, W. W.; Han, J. W. Negative Giant Surface Potential of Vacuum-Evaporated Tris(7-Propyl-8- Hydroxyquinolinolato) Aluminum(III) [Al(7-Prq)3] Film. *Org. Electron.* **2013**, *14*, 1988–1991.

(42) Noguchi, Y.; Miyazaki, Y.; Tanaka, Y.; Sato, N.; Nakayama, Y.; Schmidt, T. D.; Brütting, W.; Ishii, H. Charge Accumulation at Organic Semiconductor Interfaces Due to a Permanent Dipole Moment and Its Orientational Order in Bilayer Devices. *J. Appl. Phys.* **2012**, *7306*, 114508.

(43) Ito, E.; Washizu, Y.; Hayashi, N.; Ishii, H.; Matsuie, N.; Tsuboi, K.; Ouchi, Y.; Yamashita, K.; Seki, K. Spontaneous Buildup of Giant Surface Potential by Vacuum Deposition of Alq3 and Its Removal by Visible Light Irradiation. *J. Appl. Phys.* **2002**, *92*, 7306–7310.

(44) Baglay, R. R.; Roth, C. B. Local Glass Transition Temperature T g (z) of Polystyrene next to Different Polymers: Hard vs. Soft Confinement. *J. Chem. Phys.* **2017**, *146*, 203307.

(45) Komura, M.; Yoshitake, A.; Komiyama, H.; Iyoda, T. Control of Air-Interface-Induced Perpendicular Nanocylinder Orientation in Liquid Crystal Block Copolymer Films by a Surface-Covering Method. *Macromolecules* **2015**, *48*, 672–678.

(46) Zhang, Y.; Glor, E. C.; Li, M.; Liu, T.; Wahid, K.; Zhang, W.; Riggleman, R. A.; Fakhraai, Z. Long-Range Correlated Dynamics in Ultra-Thin Molecular Glass Films. *J. Chem. Phys.* **2016**, *145*, 114502.

(47) Mezger, M.; Jérôme, B.; Kortright, J. B.; Valvidares, M.; Gullikson, E. M.; Giglia, A.; Mahne, N.; Nannarone, S. Molecular Orientation in Soft Matter Thin Films Studied by Resonant Soft X-Ray Reflectivity. *Phys. Rev. B* **2011**, *83*, 155406.

(48) DeLongchamp, D. M.; Kline, R. J.; Fischer, D. A.; Richter, L. J.; Toney, M. F. Molecular Characterization of Organic Electronic Films. *Adv. Mater.* **2011**, *23*, 319–337.

(49) Ishii, H.; Sugiyama, K.; Ito, E.; Seki, K. Energy Level Alignment and Interfacial Electronic Structures at Organic/Metal and Organic/Organic Interfaces. *Adv. Mater.* **1999**, *11*, 605–625.

(50) Li, P.; Ingram, G.; Lee, J.-J.; Zhao, Y.; Lu, Z.-H. Energy Disorder and Energy Level Alignment between Host and Dopant in Organic Semiconductors. *Commun. Phys.* **2019**, *2*, 1–7.

(51) Kim, J. Y.; Yokoyama, D.; Adachi, C. Horizontal Orientation of Disk-like Hole Transport Molecules and Their Application for Organic Light-Emitting Diodes Requiring a Lower Driving Voltage. *J. Phys. Chem. C* **2012**, *116*, 8699–8706.

(52) Park, B.; In, I.; Gopalan, P.; Evans, P. G.; King, S.; Lyman, P. F. Enhanced Hole Mobility in Ambipolar Rubrene Thin Film Transistors on Polystyrene. *Appl. Phys. Lett.* **2008**, *92*, 112.

(53) Di, C.-A.; Liu, Y.; Yu, G.; Zhu, D. Interface Engineering: An Effective Approach toward High-Performance Organic Field-Effect Transistors. *Acc. Chem. Res.* **2009**, *42*, 1573–1583.

(54) Bakken, N.; Torres, J. M.; Li, J.; Vogt, B. D. Thickness Dependent Modulus of Vacuum Deposited Organic Molecular Glasses for Organic Electronics Applications. *Soft Matter* **2011**, *7*, 7269–7273.

(55) Torres, J. M.; Bakken, N.; Stafford, C. M.; Li, J.; Vogt, B. D. Thickness Dependence of the Elastic Modulus of Tris (8-Hydroxyquinolinato) Aluminium. *Soft Matter* **2010**, *6*, 5783–5788.

TOC graphic

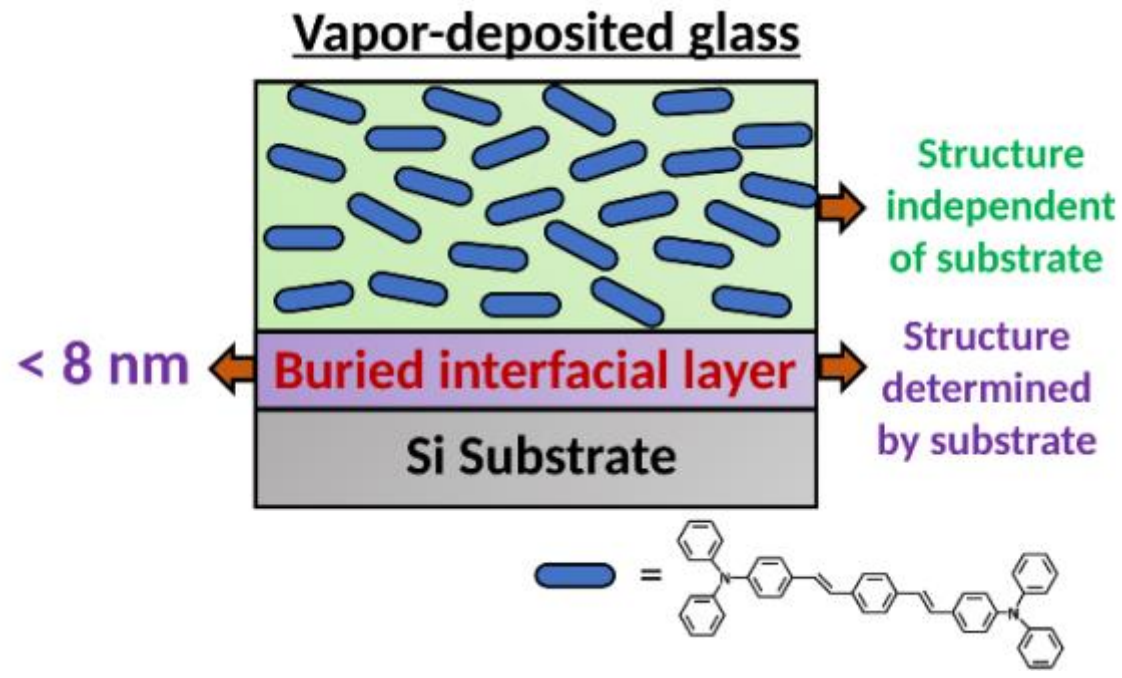
Vapor-deposited glass
Structure independent of substrate
< 8 nm
Buried interfacial layer
Structure determined by substrate
Si Substrate
=

# Supporting information

## Over what length scale does an inorganic substrate perturb the structure of a glassy organic semiconductor?

*Kushal Bagchi*[*a], Chuting Deng[b], Camille Bishop[a], Yuhui Li[c], Nicholas E Jackson[b,d], Lian Yu[a,c], M.F Toney[e], J.J de Pablo[b,d], M.D Ediger[a]*

[a]Department of Chemistry, University of Wisconsin-Madison, Madison, Wisconsin 53706, United States

[b]Pritzker School of Molecular Engineering, University of Chicago, Chicago, Illinois 60637, United States.

[c]School of Pharmacy, University of Wisconsin-Madison, 777 Highland Avenue, Madison, Wisconsin 53705-2222, United States

[d]Center for Molecular Engineering and Materials Science Division, Argonne National Laboratory, Lemont, Illinois 60439, United States, Lemont, Illinois 60439, United States

[e]Stanford Synchrotron Radiation Lightsource, SLAC National Accelerator Laboratory, Menlo Park, California 94025, United States

Corresponding author: kbagchi@wisc.edu

**(i)Morphology of ultrathin films**

Ultrathin vapor-deposited glasses often dewet upon deposition, and as a result do not form continuous films. Dewetting can influence the structure of a vapor-deposited glass. However, we are interested in deposition conditions (deposition temperature and rate) that produce continuous films upon deposition. Continuous films are desirable for device applications. To ensure we are forming continuous films we perform AFM. In figure 3 of the main manuscript we quantify an interfacial length scale for glasses deposited at 290 K. As even the thinnest film deposited at this substrate temperature does not dewet, we are confident that all the films measured for figure 3(main manuscript) form continuous films upon deposition.

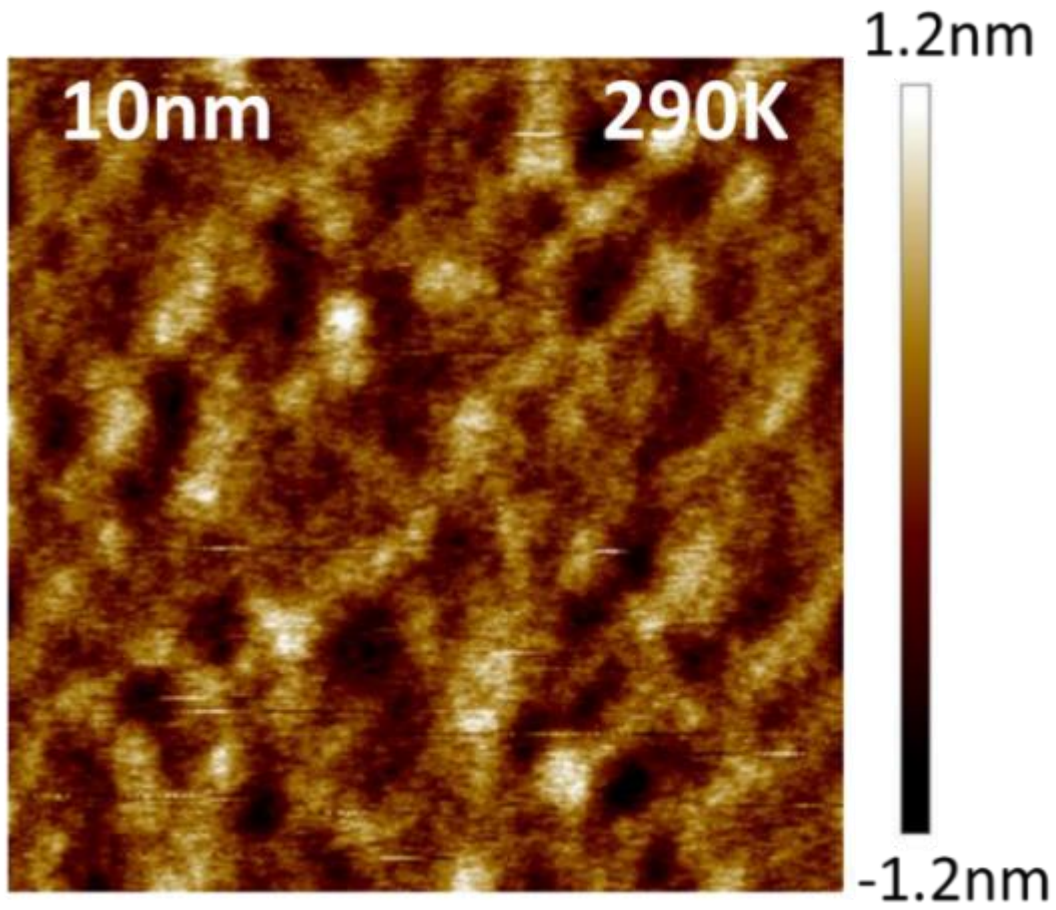

Figure S1: AFM height map of 10 nm DSA-Ph film deposited at 290 K. The AFM image shows that when a DSA-Ph glass is deposited at 290K it forms a continuous film on a $Si/SiO_2$(2nm) surface. All the data in figure 3 of the main manuscript are for films which are 10 nm or thicker and form continuous films. The scan area is 1 micron × 1 micron.

## (ii) Morphology of bare substrates

It has been hypothesized in the literature that the substrate surface roughness influences the structure of vapor-deposited glasses as thick as 100 nm. To test this hypothesis, we deposit on a smooth surface, $Si/SiO_2$(2 nm) and a rough surface, gold. We measure the substrate surface roughness of the bare substrates with AFM.

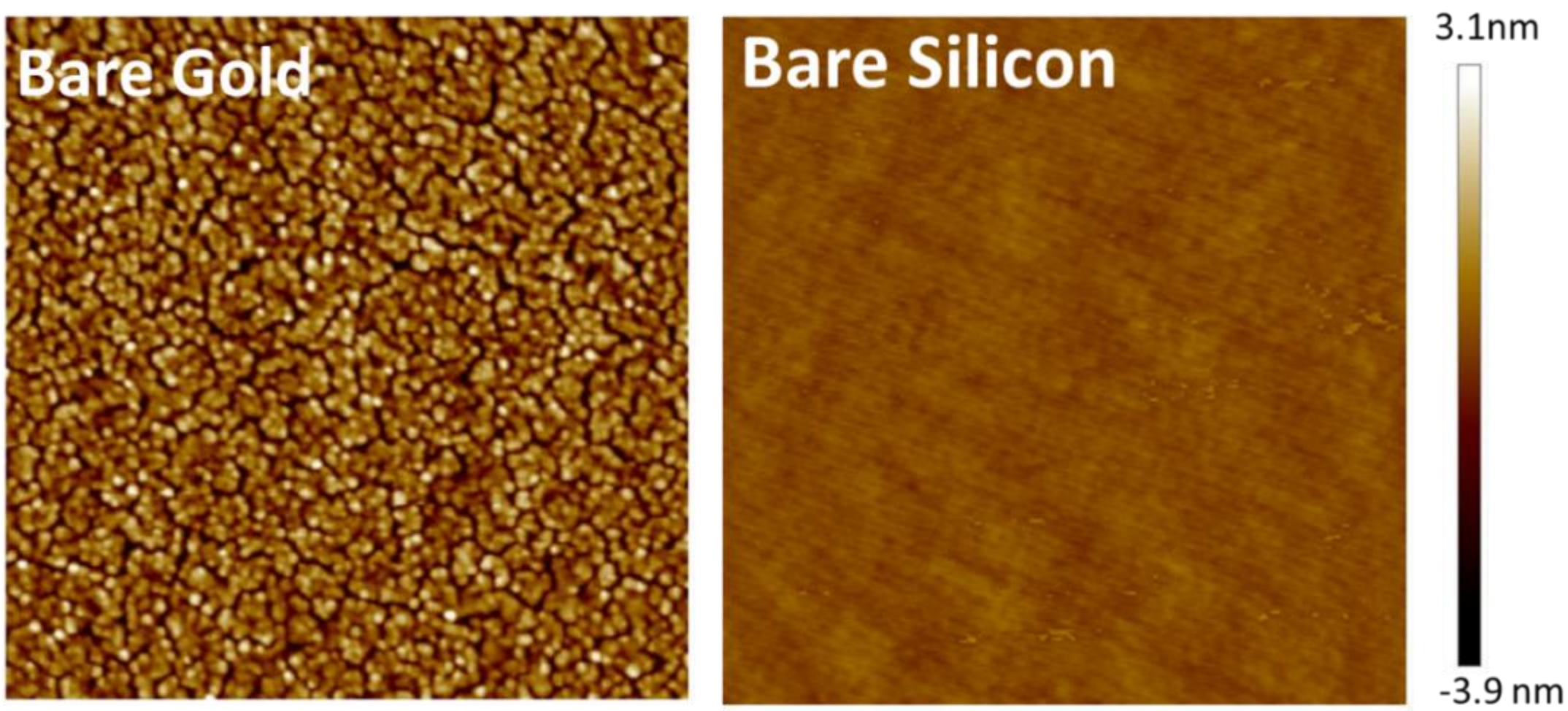

Figure S2: AFM height maps of gold and silicon (with ~ 2 nm of native oxide) substrates. The scan area is 1 micron × 1 micron.

## (iii) Interfacial length-scale of glass deposited at 252K

In Figure 3 of the main manuscript we show the GIWAXS order parameter as a function of film thickness at a deposition temperature of 290 K. Shown in Figure S3 is the analogous plot for glasses deposited at 252 K. Figure S3 shows a similar trend with film thickness is seen at a deposition temperature of 252 K.

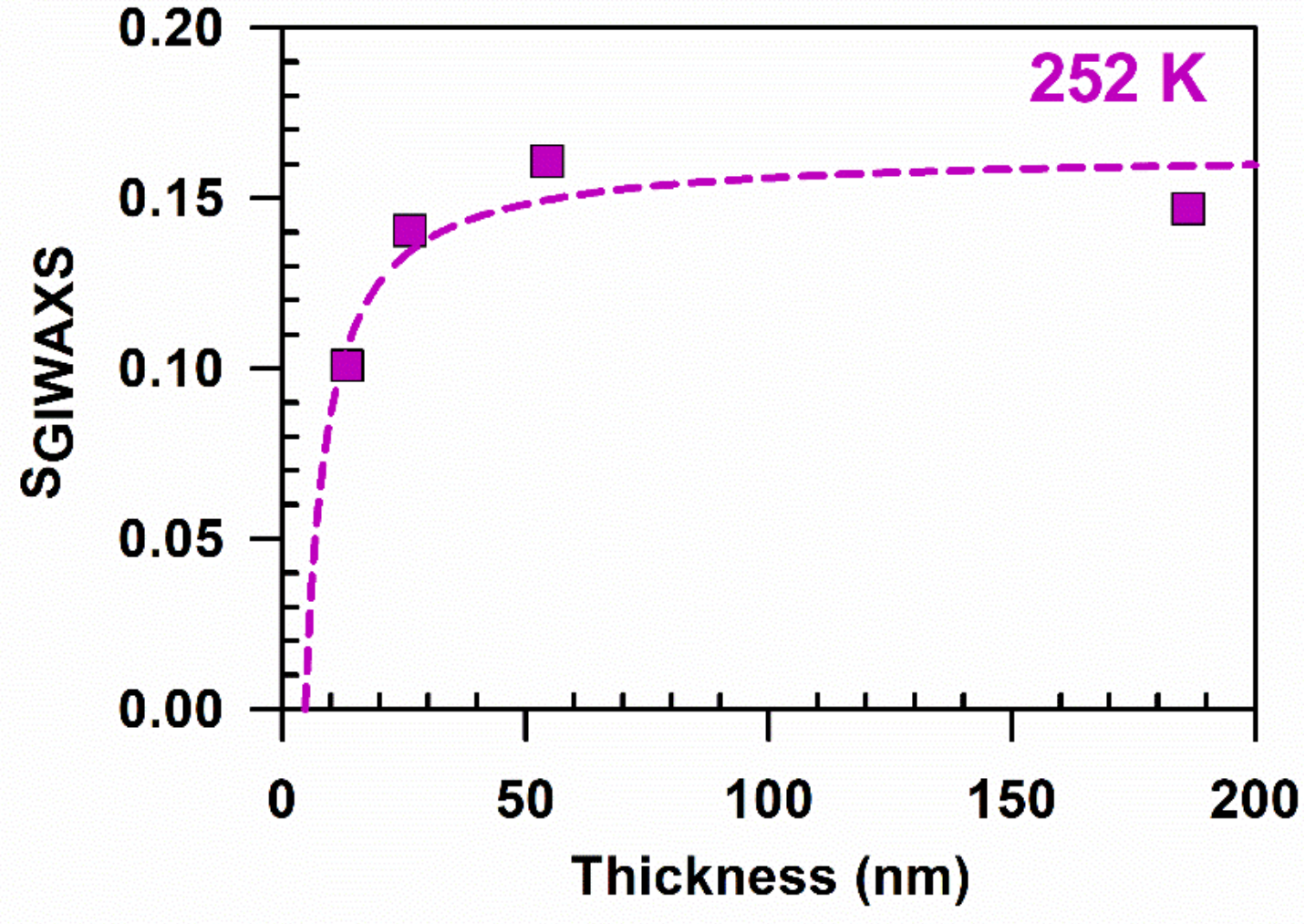

Figure S3: The $S_{GIWAXS}$ order parameter is plotted as a function of film thickness for DSA-Ph glasses prepared at 252 K. The interfacial length scale obtained from the two-layer fit is 4.7 ± 1.6 nm. Samples were deposited on a silicon substrate with ~2nm of native oxide.

## (iv) GIWAXS diffraction patterns

In Figure 2 and 3 of the main manuscript the GIWAXS order parameter, $S_{GIWAXS}$, is evaluated for films of different thickness deposited across a range of substrate temperatures. While the GIWAXS order parameter is a quantitative measure, its calculation requires significant data processing. The unprocessed diffraction patterns provide direct, albeit qualitative, information about molecular packing. The unprocessed diffraction patterns provide direct evidence for the conclusions in figure 2 and 3.

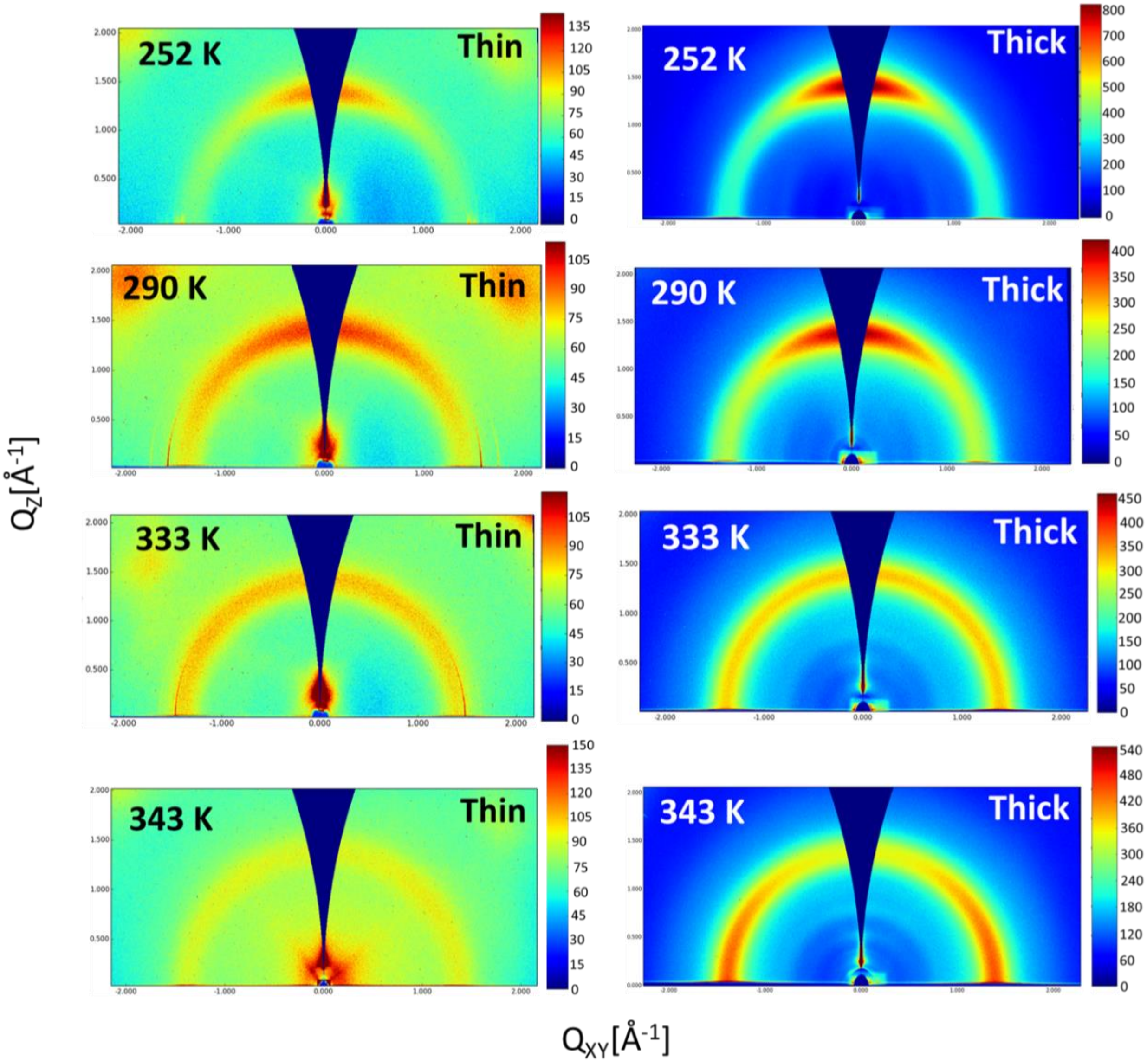

Figure S4: GIWAXS diffraction patterns from DSA-Ph glasses deposited at different substrate temperatures. The $S_{GIWAXS}$ order parameter for these glasses are shown in figure 2 of the main text. In this figure, the “thin” samples are ~25 nm in thickness and the “thick” samples are ~175 nm in thickness.

## (v) Structure at the surface of vapor-deposited glasses.

To evaluate how much the free surface contributes to the lowering of order observed for the thinnest films in Figure 3 of the main text and Figure S3 of the supplemental text, we determine the $S_{GIWAXS}$ order parameter for incidence angles below the critical angle. An order parameter evaluated at an incidence angle below the critical angle is sensitive to the top ~5 nm of the film. We observe that the free surface exhibits similar packing to the bulk at all substrate temperatures.

At $T_{sub}$=252 K and $T_{sub}$=290 K, however the free surface is slightly less ordered which means that the free surface also contributes slightly to the lowering of order observed in the thinnest films.

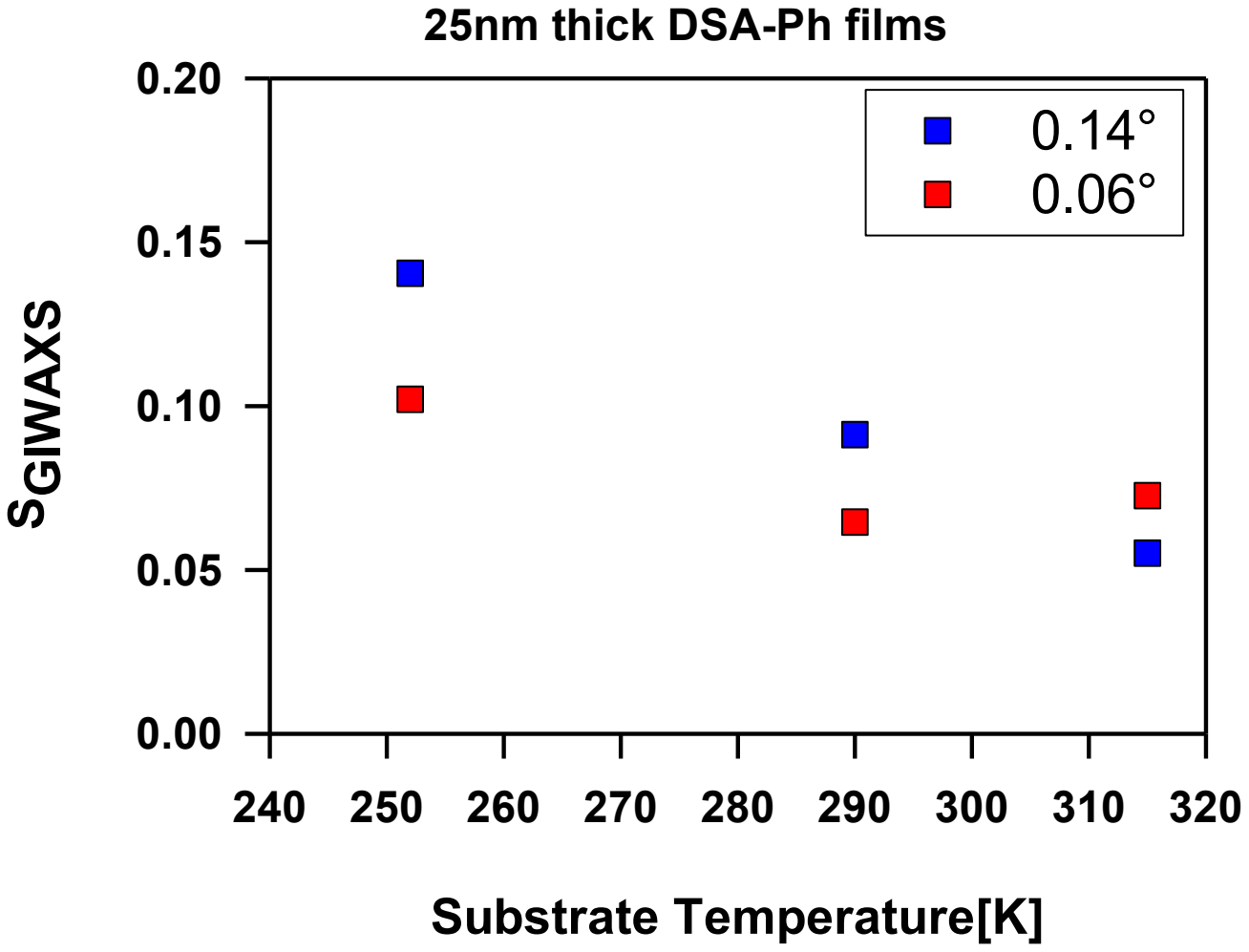

Figure S5: The $S_{GIWAXS}$ order parameter at incidence angles below (0.06°) and above (0.14°) the critical angle. Data above the critical angle is a measure of the average bulk structure while data below the critical angle is a measure of the structure in the top ~ 5 nm.

## (vi) Scattering lineshapes

The peak position and peak width of the "amorphous halo" peak carry important structural information about a glass. We observe that the scattering lineshapes, for films of 18 nm and 604 nm, both deposited at 290 K are very similar. This provides further support for the similarity in packing in these two glasses that vary in thickness by a factor of ~ 30.

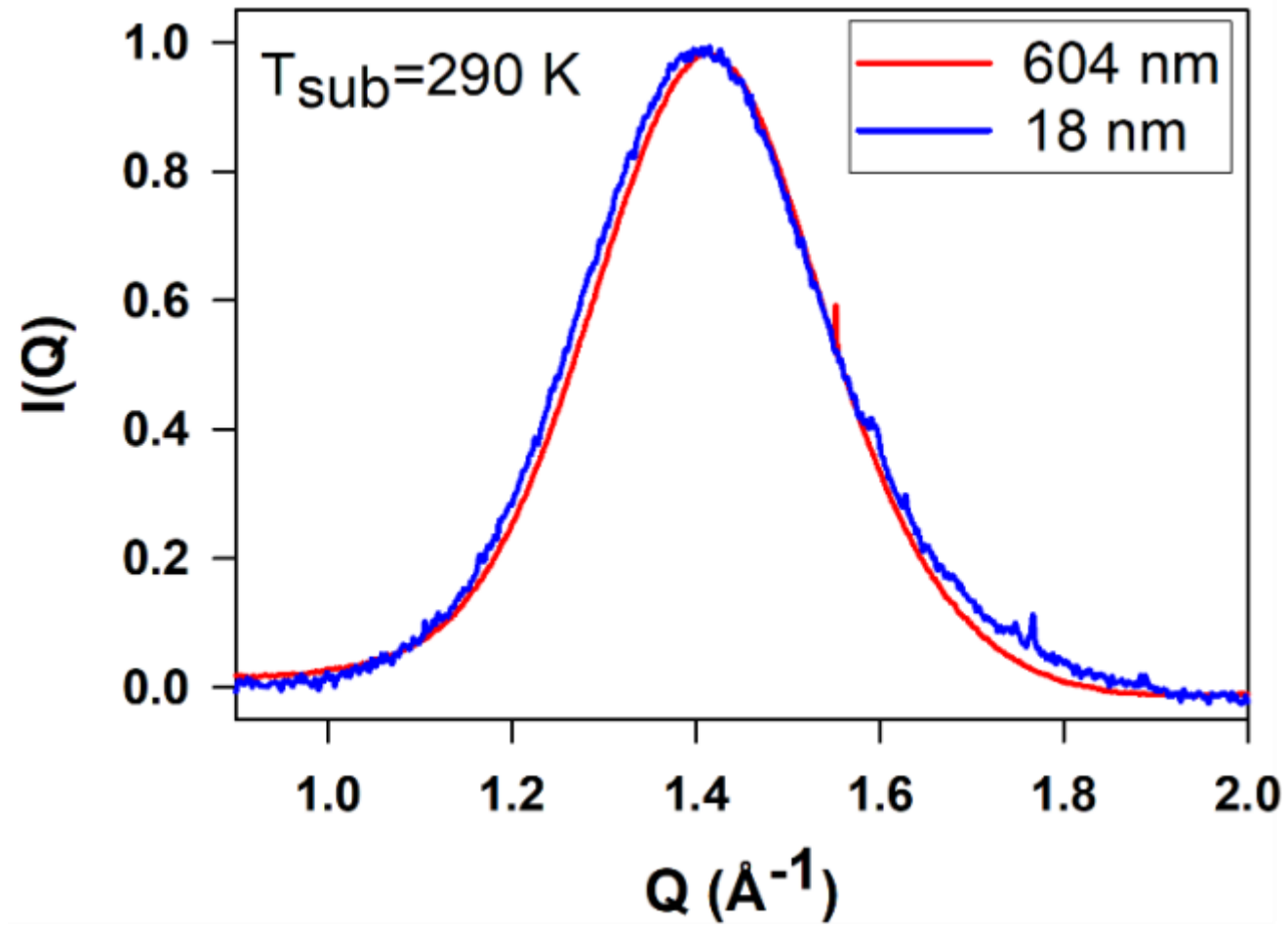

Figure S6: The scattering intensity, averaged over all accessible angles in reciprocal space, as function of Q for the two patterns shown in Figure 1 of the main text. The data has been background subtracted and normalized.

## (vii) Simulation Box

Shown in Fig S7 is a snapshot of the simulation box used in our simulation.

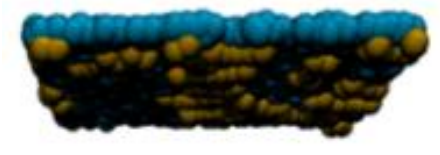

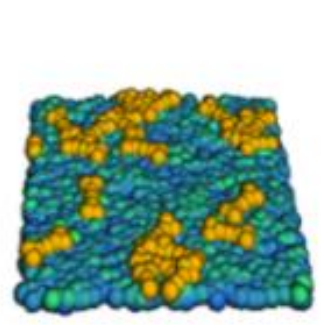

Figure S7: Schematic of the simulation box. A molecule is inserted in the middle of the box and diffuses to a substrate either at the top or at the bottom. The box dimension is 28σ x 28σ x 130σ. The box is periodic in x and y direction (substrate plane) and nonperiodic in z direction (substrate normal). Substrate atoms are colored based on their position in z-direction (substrate normal).

In order to control molecular orientation at the buried interface (in simulations), we define Lennard- Jones parameters between the substrate beads and the beads of the molecules. Shown in Table S1 are the Lennard-Jones parameters for the two different interaction potentials between the molecule and the substrate in our simulations.

A)

| Pair | | Non-anchoring | | Vertically anchoring | |
|---|---|---|---|---|---|
| | | ε | σ | ε | σ |
| 1 | 4 | 1.0 | 1.0 | 5.0 | 1.0 |
| 2 | 4 | 1.0 | 1.0 | 0.2 | 1.4 |
| 3 | 4 | 1.0 | 1.0 | 0.2 | 2.5 |

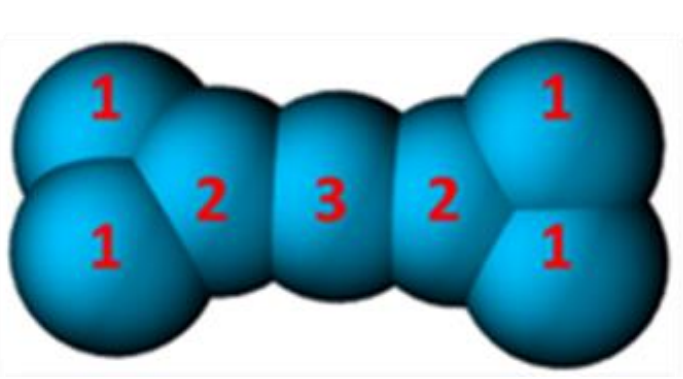

B)

| Pair | | Non-anchoring/vertically anchoring | |
|---|---|---|---|
| | | ε | σ |
| 1 | 1 | 1.0 | 1.0 |
| 1 | 2 | 1.0 | 1.0 |
| 1 | 3 | 1.0 | 1.0 |
| 2 | 2 | 1.0 | 1.0 |
| 2 | 3 | 1.0 | 1.0 |
| 3 | 3 | 1.0 | 1.0 |
| 4 | 4 | 0.1 | 0.6 |

Table S1: Lennard-Jones parameters for interfacial potentials (A) and inter-bead potentials (B). Atom types 1-3 are labelled on the coarse-grained molecule. Type 4 atoms are substrate atoms.

## (viii) Determination of fictive temperature in simulations

To characterize the glass transition behavior of the coarse-grained DSA-Ph molecule, the equilibrium liquid at high temperature (T=1.5) is linearly cooled to a low temperature (T=0.1) at four different cooling rates ranging from $q_c=10^{-3}$ to $q_c=10^{-6}$ (temperature/time in dimensionless units). Linear regressions are performed between density and temperature for the liquid regime and glass regime respectively. The fictive temperature is defined as the intersection of the linear extrapolations from the two regimes.

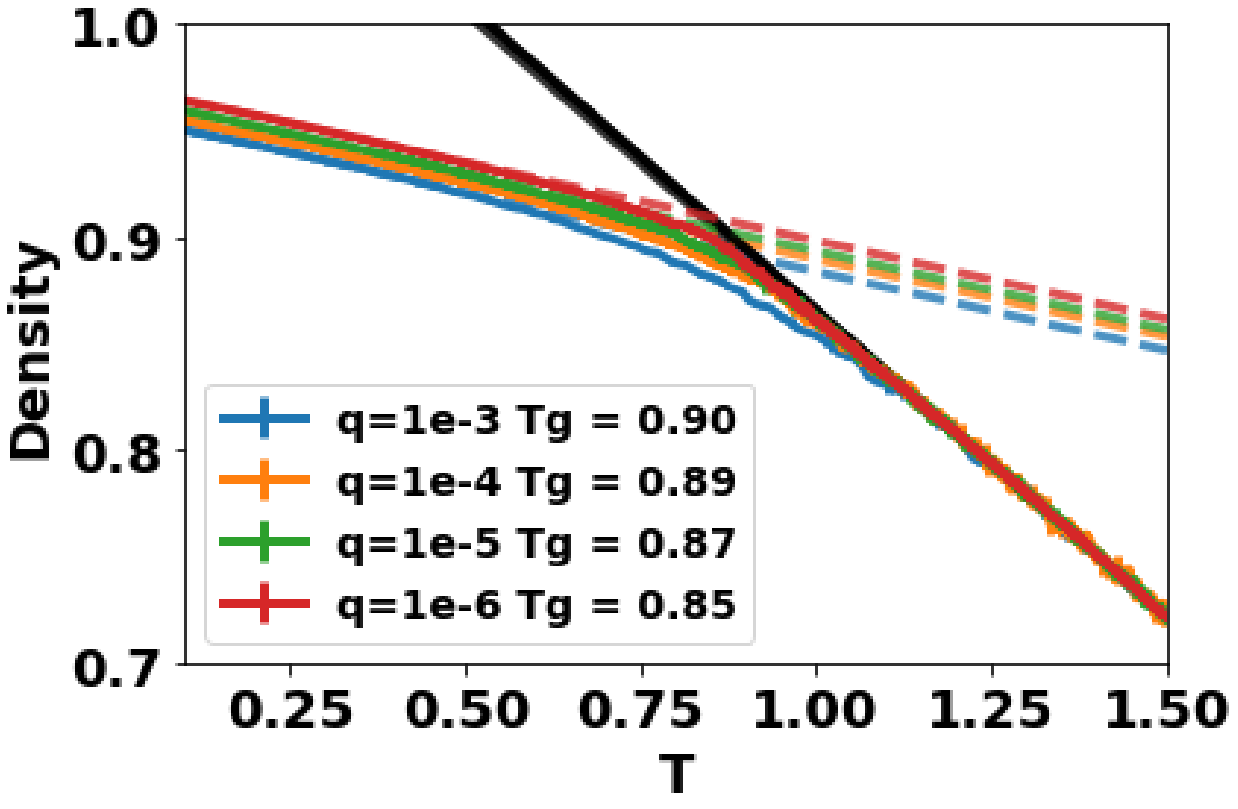

Figure S8: Bulk density as a function of temperature during cooling. The cooling rates are q=$10^{-3}$, $10^{-4}$, $10^{-5}$, and $10^{-6}$. The fictive temperature determined at the slowest cooling rate is 0.85.

## (ix) Characterization of molecular orientation near the substrate for different interaction potentials (in simulations)

From a trajectory where 80 molecules were deposited onto a substrate, a histogram of $|\cos\theta|$ is collected to characterize the surface orientation, where $\theta$ is the angle between the longitudinal axis of the molecule and the substrate normal, and $|\ldots|$ denotes absolute value. This calculation is performed at a substrate temperature of 0.68.

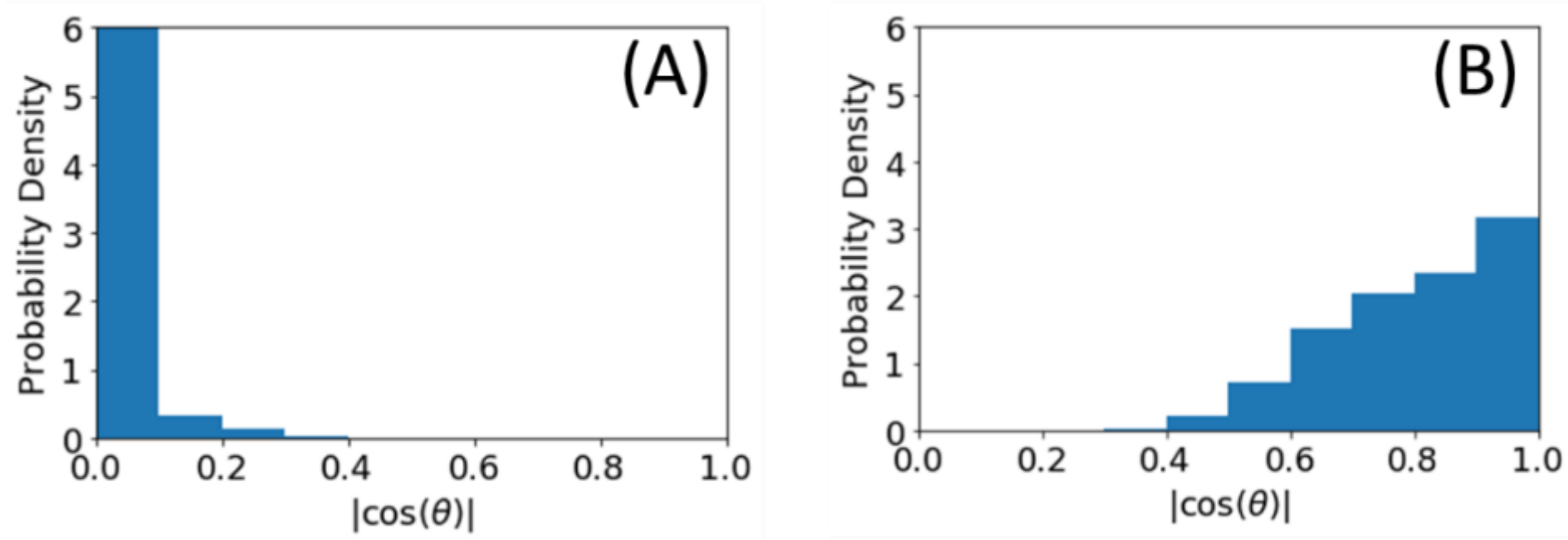

Figure S9: Distribution of molecule orientation at the substrate surface with non-anchoring potential (A) and with vertically anchoring potential (B). The average order parameters, $P_2$, are -0.50 for non-anchoring potential (A) and 0.52 for vertically anchoring potential (B), respectively.